\documentclass[11pt]{article}
\usepackage{graphicx}
\usepackage{amsmath}
\usepackage{amssymb}
\usepackage{caption2}
\begin{document}
\bibliographystyle{prsty}
\begin{center}
{\large {\bf \sc{  Analysis of the pseudoscalar partner of the $Y(4660)$ and related bound states}}} \\[2mm]
Zhi-Gang Wang \footnote{wangzgyiti@yahoo.com.cn.  }, Xiao-Hong Zhang     \\
 Department of Physics, North China Electric Power University,
Baoding 071003, P. R. China
\end{center}

\begin{abstract}
In this article, we study the pseudoscalar   bound state
$\eta_c'f_0(980)$ (irrespective of the hadro-charmonium and the
molecular state) with the QCD sum rules. Considering the $SU(3)$
symmetry of the light flavor quarks and the heavy quark symmetry, we
also study the bound states $\eta_c'\sigma(400-1200)$,
$\eta_b'''f_0(980)$ and $\eta_b'''\sigma(400-1200)$, and make
reasonable predictions for their masses.
\end{abstract}



\section{Introduction}
  In 2007, the Belle collaboration observed two resonant structures  in the $\pi^+\pi^-
\psi'$ invariant mass distribution in the  cross section for the
process $e^+ e^- \to \pi^+\pi^- \psi'$ between threshold and
$\sqrt{s}=5.5~\rm{GeV}$
 using $673~\rm{fb}^{-1}$ of data on and off the
$\Upsilon(4S)$ ($\Upsilon'''$) resonance, one at  $4361\pm 9\pm
9~\rm{MeV}$ with a width of $74\pm 15\pm 10~\rm{MeV}$, and another
at $4664\pm 11\pm 5~\rm{MeV}$ with a width of $48\pm 15\pm
3~\rm{MeV}$ (they are denoted as the $Y (4360)$ and $Y(4660)$
respectively) \cite{Yexp}. The structure $Y(4660)$ is neither
observed in the initial state radiation (ISR) process
$e^+e^-\to\gamma_{ISR}\pi^+\pi^-J/\psi$ \cite{ISRJpsi}, nor in the
exclusive cross processes  $e^+e^-\to D{\bar D},D{\bar D}^*,
D^*{\bar D}^*, D{\bar D}\pi$, $ J/\psi D^{(*)} {\bar D}^{(*)}$
\cite{ExpDDbar1,ExpDDbar2,ExpDDbar3,ExpDDbar4,Abe:2007sy}.

There have been several  canonical charmonium interpretations for
the $Y(4660)$, such as the $5^3S_1$ state \cite{Ding0708}, the
$6^3S_1$ state \cite{Chao0903}, the $5^3S_1-4^3D_1$ mixing state
\cite{SDmix}, and some exotic  interpretations,  such as the radial
excited state of the
$\frac{1}{\sqrt{2}}(|\Lambda_c\bar{\Lambda}_c\rangle+|\Sigma^0_c\bar{\Sigma}^0_c\rangle)$
\cite{Qiao0709}, the  vector $cs\bar{c}\bar{s}$ tetraquark state
\cite{Nielsen0804}, etc.

A critical information for understanding the structure of the
charmonium-like states is wether or not the $\pi\pi$ comes from a
resonance. There is some indication that only the $Y(4660)$ has a
well defined intermediate state which is consistent with the scalar
meson $f_0(980)$ in the $\pi\pi$ invariant mass spectra
\cite{babarconf0801}.

 In Refs.\cite{Voloshin0803,VoloshinReV},
Voloshin et al argue that the charmonium-like states $Y(4660)$,
$Z(4430)$, $Y(4260)$, $\cdots$ may be hadro-charmonia. The
relatively compact charmonium states ($J/\psi$, $\psi'$ and
$\chi_{cJ}$) can be bound inside light hadronic matter, in
particular inside higher resonances made from light quarks and (or)
gluons. The charmonium state in such binding
 retains its properties essentially,  the bound system
(hadro-charmonium, a special molecular state) decays into light
mesons and the particular charmonium.

In Ref.\cite{Guo0803}, Guo et al assume that the $Y(4660)$ is a
$\psi'f_0(980)$ bound state (molecular state), as  the nominal
threshold of the $\psi'-f_0(980)$ system is about $4666\pm10$~MeV
\cite{PDG}, the $Y(4660)$  decays  dominantly via the decay of the
scalar meson $f_0(980)$, $Y(4660)\to\psi'f_0(980)\to \psi'\pi\pi$, $
\psi'K{\bar K}$, the difficulties in the canonical charmonium
interpretation is overcome. Considering the heavy quark spin
symmetry, Guo et al predict an $\eta_c'f_0(980)$ bound state
($Y(4616)$) as the spin-doublet partner of the Y(4660)  with a mass
of $4616^{+5}_{-6}\,\rm{MeV}$ and a width of $60\pm30 \,\rm{MeV}$
through the prominent decay mode
   $ \eta_c'\pi \pi$ \cite{GuoPRL}.

In previous work \cite{WangZhang}, we studied the mass of the
$Y(4660)$ as a $\psi'f_0(980)$ bound state (irrespective of the
hadro-charmonium and the molecular state) using the QCD sum rules
\cite{SVZ79,Reinders85}. In this article, we extend our previous
work to study the $\eta_c'f_0(980)$ bound state  $Y(4616)$,
furthermore, we take into account the $SU(3)$ symmetry of the light
flavor quarks and the heavy quark symmetry, study the related hidden
charm and hidden bottom states. In the QCD sum rules, the operator
product expansion is used to expand the time-ordered currents into a
series of quark and gluon condensates which parameterize the long
distance properties of the QCD vacuum. Based on the quark-hadron
duality, we can obtain copious information about the hadronic
parameters at the phenomenological side \cite{SVZ79,Reinders85}.

The article is arranged as follows:  we derive the QCD sum rules for
  the  pseudoscalar charmonium-like state $Y(4616)$ and the related bound  states
  in section 2; in section 3, numerical
results and discussions; section 4 is reserved for conclusion.

\section{QCD sum rules for  the $Y(4616)$ and related bound  states }
In the following, we write down  the two-point correlation functions
$\Pi(p)$  in the QCD sum rules,
\begin{eqnarray}
\Pi(p)&=&i\int d^4x e^{ip \cdot x} \langle
0|T\left[J/\eta(x)J/\eta^{\dagger}(0)\right]|0\rangle \, , \\
J(x)&=&\bar{Q}(x)i\gamma_5 Q(x) \bar{s}(x)s(x) \, , \nonumber\\
\eta(x)&=&\frac{1}{\sqrt{2}}\bar{Q}(x)i\gamma_5 Q(x)
\left[\bar{u}(x)u(x)+\bar{d}(x)d(x) \right]\, ,
\end{eqnarray}
where the $Q$ denotes the heavy quarks $c$ and $b$. We use the
currents $J(x)$ and $\eta(x)$ ($Q=c$) to interpolate the bound
states $\eta_c'f_0(980)$ (predicted in Ref.\cite{GuoPRL}) and
$\eta_c'\sigma(400-1200)$, respectively.

The hidden charm current $\bar{c}(x)i\gamma_5 c(x)$ can interpolate
the charmonia $\eta_c$, $\eta_c'$, $\eta_c''$,  $\cdots$; and the
hidden bottom current $\bar{b}(x)i\gamma_5 b(x)$ can interpolate the
bottomonia $\eta_b$, $\eta_b'$, $\eta_b''$,  $\cdots$. We assume
that the scalar mesons $f_0(980)$ and $\sigma(400-1200)$ are the
conventional $q\bar{q}$ states, to be more precise, they have large
$q\bar{q}$ components, while in Refs.\cite{Guo0803,GuoPRL}  the
scalar meson  $f_0(980)$ is taken as the $K\bar{K}$ molecular state.
There are  hot controversies  about their nature, for example, the
conventional $q\bar{q}$ states (strongly affected  by the nearby
thresholds) \cite{Wang2004}, the tetraquark states, the molecular
states \cite{Close2002,ReviewScalar}. The currents $J(x)$ and
$\eta(x)$ ($Q=c$) have non-vanishing couplings with the bound states
$\eta_c f_0(980)$, $\eta_c'f_0(980)$, $\eta_c''f_0(980)$, $\cdots$
and $\eta_c\sigma(400-1200)$, $\eta_c'\sigma(400-1200)$,
$\eta_c''\sigma(400-1200)$, $\cdots$, respectively. Considering the
heavy quark symmetry, there maybe exist some hidden bottom  bound
states, for example, $\eta_b f_0(980)$, $\eta_b' f_0(980)$,
$\eta_b'' f_0(980)$, $\eta_b''' f_0(980)$, $\eta_b\sigma(400-1200)$,
$\eta_b'\sigma(400-1200)$, $\eta_b''\sigma(400-1200)$,
$\eta_b'''\sigma(400-1200)$, $\cdots$, we study those possibilities
with the  currents $J(x)$ and $\eta(x)$ ($Q=b$), and make
predictions for their masses.

We can insert  a complete set of intermediate hadronic states with
the same quantum numbers as the current operators $J(x)$ and
$\eta(x)$ into the correlation functions  $\Pi(p)$  to obtain the
hadronic representation \cite{SVZ79,Reinders85}. After isolating the
ground state contributions  from the pole terms of the $Y$, we get
the following result,
\begin{eqnarray}
\Pi(p)&=&\frac{\lambda_{Y}^2}{M_{Y}^2-p^2} +\cdots \, \, ,
\end{eqnarray}
where the pole residues (or couplings) $\lambda_Y$ are defined by
\begin{eqnarray}
\lambda_{Y} &=& \langle 0|J/\eta(0)|Y(p)\rangle  \, .
\end{eqnarray}

The  contributions from the two-particle and many-particle reducible
states are  small enough to be neglected \cite{Lee2005}, for
example,
\begin{eqnarray}
\Pi_{2}(p)&=&i\lambda_{\eta_c f_0}^2\int~{d^4q\over(2\pi)^4}
\frac{1}{\left[ q^2-m_{\eta_c}^2\right]\left[
(p-q)^2-m_{f_0}^2\right]}+\cdots \, ,
\end{eqnarray}
where the pole residue (or coupling) $\lambda_{\eta_c f_0}$ is
defined by
\begin{eqnarray}
\langle0|J(0)|\eta_c f_0(p)\rangle=\lambda_{\eta_c f_0}\, .
\end{eqnarray}
The  coupling $\lambda_{\eta_c f_0}$ can be written in terms of the
$\eta_c$ meson decay constant $f_{\eta_c}$ and the coupling
$\lambda_{f_0}$ of the scalar meson $f_0(980)$  with a  tetraquark
current. The coupling $\lambda_{f_0}$  should be very small as the
$f_0(980)$ is a light flavor meson, and the two-particle reducible
contributions  can be neglected \cite{Lee2005,Nielsen0907}.
Furthermore, the pseudoscalar charmonia $\eta_c$, $\eta_c'$,
$\eta_c''$, $\cdots$ and the pseudoscalar bottomonia $\eta_b$,
$\eta_b'$, $\eta_b''$, $\cdots$
 also have Fock states with additional $q\bar{q}$ components beside
 the $Q\bar{Q}$ components. The currents $J(x)$ and
 $\eta(x)$ may have non-vanishing couplings  with the pseudoscalar charmonia
 and pseudoscalar bottomonia, those couplings are supposed to be small, as the
 main Fock states of the pseudoscalar charmonia
 and pseudoscalar bottomonia are the $Q\bar{Q}$ components, and the  pseudoscalar charmonia
 and pseudoscalar bottomonia have much smaller masses than the corresponding
 molecular states $Y$.

 After performing the standard procedure of the QCD sum rules, we obtain   two  sum rules in the
 $c\bar{c}s\bar{s}$ and $b\bar{b}s\bar{s}$ channels respectively:
\begin{eqnarray}
\lambda_{Y}^2 e^{-\frac{M_Y^2}{M^2}}= \int_{\Delta}^{s_0} ds
\rho(s)e^{-\frac{s}{M^2}} \, ,
\end{eqnarray}
\begin{eqnarray}
\rho(s)&=&\rho_{0}(s)+\rho_{\langle
\bar{s}s\rangle}(s)+\rho_{\langle GG\rangle}(s)\langle
\frac{\alpha_s GG}{\pi}\rangle+\rho_{\langle
\bar{s}s\rangle^2}(s)+\rho_{\langle GGG\rangle}(s)\langle
g_s^3f_{abc}G^aG^bG^c \rangle\, .
\end{eqnarray}
The explicit expressions of the spectral densities $\rho_{0}(s)$,
$\rho_{\langle \bar{s}s\rangle}(s)$, $\rho_{\langle GG\rangle}(s)$,
 $\rho_{\langle \bar{s}s\rangle^2}(s)$, and $\rho_{\langle GGG\rangle}(s)$ are presented in  the
appendix.   The $s_0$ is the continuum threshold parameter and the
$M^2$ is the Borel
      parameter.  We can obtain  two sum rules in  the
$c\bar{c}q\bar{q}$ and $b\bar{b}q\bar{q}$ channels with a simple
replacement $m_s\rightarrow m_q$,  $\langle
\bar{s}s\rangle\rightarrow\langle \bar{q}q\rangle$ and $\langle
\bar{s}g_s \sigma Gs\rangle\rightarrow\langle \bar{q}g_s \sigma
Gq\rangle$.

 We carry out the operator
product expansion (OPE) to the vacuum condensates adding up to
dimension-10. In calculation, we
 take  assumption of vacuum saturation for  high
dimension vacuum condensates, they  are always
 factorized to lower condensates with vacuum saturation in the QCD sum rules,
  factorization works well in  the large $N_c$ limit. In this article, we take into account the contributions from the
quark condensates $\langle \bar{s}s\rangle$, $\langle
\bar{s}s\rangle^2$,  mixed condensates $\langle \bar{s}g_s\sigma
Gs\rangle$, $\langle \bar{s}s\rangle\langle \bar{s}g_s\sigma
Gs\rangle$, $\langle \bar{s}g_s\sigma Gs\rangle^2$ and   the
  gluon condensates $\langle \frac{\alpha_sGG}{\pi}\rangle$, $\langle
g_s^3f_{abc}G^aG^bG^c \rangle $ (one can see the appendix for the
explicit expressions). The contributions from the quark-gluon
condensates $\langle \frac{\alpha_sGG}{\pi}\rangle
\langle\bar{s}s\rangle$, $\langle \frac{\alpha_sGG}{\pi}\rangle
\langle\bar{s}g_s\sigma Gs\rangle$, $\langle g_s^3f_{abc}G^aG^bG^c
\rangle\langle\bar{s}s\rangle $, $\langle
\frac{\alpha_sGG}{\pi}\rangle \langle\bar{s}s\rangle^2$  are
suppressed by large denominators and would not play any significant
roles. Comparing with the gluon condensate $\langle
\frac{\alpha_sGG}{\pi}\rangle$, the vacuum condensates $\langle
g_s^3f_{abc}G^aG^bG^c \rangle $, $\langle
\frac{\alpha_sGG}{\pi}\rangle \langle\bar{s}g_s\sigma Gs\rangle$,
$\langle g_s^3f_{abc}G^aG^bG^c \rangle\langle\bar{s}s\rangle $ are
of
 higher  order in $\frac{\alpha_s}{\pi}$, their contributions are
 greatly  suppressed. In calculation, we observe that the contributions from the term $\langle g_s^3f_{abc}G^aG^bG^c \rangle $
 are less  than  $0.2\%$, and can be neglected safely.

 Differentiate  the Eq.(7) with respect to  $\frac{1}{M^2}$, then eliminate the
 pole residue $\lambda_{Y}$, we can obtain the  sum rule for
 the mass  of the bound state $Y$,
 \begin{eqnarray}
 M_Y^2= \frac{\int_{\Delta}^{s_0} ds
\frac{d}{d(-1/M^2)}\rho(s)e^{-\frac{s}{M^2}} }{\int_{\Delta}^{s_0}
ds \rho(s)e^{-\frac{s}{M^2}}}\, .
\end{eqnarray}

\section{Numerical results and discussions}
The input parameters are taken to be the standard values $\langle
\bar{q}q \rangle=-(0.24\pm 0.01 \,\rm{GeV})^3$, $\langle \bar{s}s
\rangle=(0.8\pm 0.2 )\langle \bar{q}q \rangle$, $\langle
\bar{q}g_s\sigma Gq \rangle=m_0^2\langle \bar{q}q \rangle$, $\langle
\bar{s}g_s\sigma Gs \rangle=m_0^2\langle \bar{s}s \rangle$,
$m_0^2=(0.8 \pm 0.2)\,\rm{GeV}^2$, $\langle \frac{\alpha_s
GG}{\pi}\rangle=(0.33\,\rm{GeV})^4 $, $\langle g_s^3f_{abc}G^a G^b
G^c \rangle=0.045\,\rm{GeV}^6$, $m_u=m_d\approx 0$,
$m_s=(0.14\pm0.01)\,\rm{GeV}$, $m_c=(1.35\pm0.10)\,\rm{GeV}$ and
$m_b=(4.8\pm0.1)\,\rm{GeV}$ at the energy scale   $\mu=1\, \rm{GeV}$
\cite{SVZ79,Reinders85,Ioffe2005}.

In the conventional QCD sum rules \cite{SVZ79,Reinders85}, there are
two criteria (pole dominance and convergence of the operator product
expansion) for choosing  the Borel parameter $M^2$ and threshold
parameter $s_0$.  We impose the two criteria on the pseudoscalar
charmonium-like states $Y$ to choose the Borel parameter $M^2$ and
threshold parameter $s_0$.

We  take the threshold parameter of the pseudoscalar
 bound state $\eta_c'f_0(980)$  as
$s^0_{s\bar{s}}=(4.62+0.5)^2\, \rm{GeV}^2\approx 26 \, \rm{GeV}^2$
tentatively to take into account possible contribution from the
ground state, where  the energy gap between the ground state and the
first radial excited state is chosen to be $0.5\,\rm{GeV}$. Taking
into account the $SU(3)$ symmetry of the light flavor quarks, we
expect the threshold parameter $s^0_{q\bar{q}}$ (for the bound state
$\eta_c'\sigma(400-1200)$) is slightly smaller than  the
$s^0_{s\bar{s}}$. Furthermore, we take into account the mass
difference between the $c$ and $b$ quarks, the threshold parameters
in the   hidden bottom channels are tentatively taken as
$s^0_{q\bar{q}}=142\, \rm{GeV}^2$ and $s^0_{s\bar{s}}=144\,
\rm{GeV}^2$. In this article,  we  use those  value as a guide to
determine the threshold parameters  $s_0$ with the QCD sum rules.

The contributions from the high dimension vacuum condensates  in the
operator product expansion are shown in Figs.1-2, where (and
thereafter) we  use the $\langle\bar{q}q\rangle$ to denote the quark
condensates $\langle\bar{q}q\rangle$, $\langle\bar{s}s\rangle$ and
the $\langle\bar{q}g_s \sigma Gq\rangle$ to denote the mixed
condensates $\langle\bar{q}g_s \sigma Gq\rangle$, $\langle\bar{s}g_s
\sigma Gs\rangle$. From the figures, we can see that the
contributions from the high dimension condensates change  quickly
with variation of the Borel parameter at the values $M^2\leq
2.7\,\rm{GeV}^2$ and $M^2\leq 7.6\,\rm{GeV}^2$ in the hidden charm
 and hidden  bottom channels respectively, such an unstable behavior
cannot lead to stable sum rules, our numerical results confirm this
conjecture, see Fig.4.

At the values $M^2\geq 2.7\,\rm{GeV}^2$ and $s_0\geq
26\,\rm{GeV}^2$, the contributions from the  $\langle
\bar{q}q\rangle^2+\langle \bar{q}q\rangle \langle \bar{q}g_s \sigma
Gq\rangle $ term are less than  $19.5\%$ in the $c\bar{c}s\bar{s}$
channel, the corresponding contributions are less than  $ 35.5\%$ in
the $c\bar{c}q\bar{q}$  channel; the contributions from the vacuum
condensate of the highest dimension $\langle\bar{q}g_s \sigma
Gq\rangle^2$ are less than $7\%$ in all the hidden charm channels,
we expect the operator product expansion is convergent in the hidden
charm channels. At the values $M^2\geq 7.6\,\rm{GeV}^2$ and $s_0\geq
144\,\rm{GeV}^2$, the contributions from the $\langle
\bar{q}q\rangle^2+\langle \bar{q}q\rangle \langle \bar{q}g_s \sigma
Gq\rangle $ term are less than  $7\%$ in the $b\bar{b}s\bar{s}$
channel, the corresponding contributions are less than  $ 18\%$ in
the $b\bar{b}q\bar{q}$  channel; the contributions from the vacuum
condensate of the highest dimension $\langle\bar{q}g_s \sigma
Gq\rangle^2$ are less than (or equal) $6\%$ in all the hidden bottom
channels, we expect the operator product expansion is convergent in
the hidden bottom channels.

The contributions from the  gluon condensate $\langle \frac{\alpha_s
GG}{\pi}\rangle$ are less than (or equal) $37\%$ ($26\%$) in the
$c\bar{c}s\bar{s}$ ($c\bar{c}q\bar{q}$) channel at the values
$M^2\geq 2.7\,\rm{GeV}^2$ and $s_0\geq 26\,\rm{GeV}^2$; while the
contributions  are less than $19.5\%$ ($16.5\%$) in the
$b\bar{b}s\bar{s}$ ($b\bar{b}q\bar{q}$) channel at the values
$M^2\geq 7.6\,\rm{GeV}^2$ and $s_0\geq 144\,\rm{GeV}^2$. The
contributions from the high dimension condensates $\langle
\frac{\alpha_s GG}{\pi} \rangle \left[\langle \bar{q} q \rangle
+\langle \bar{q}g_s\sigma G q \rangle+ \langle \bar{q} q
\rangle^2\right]$ are small enough and neglected safely.

In the QCD sum rules for the tetraquark states (irrespective of the
molecule type and the diquark-antidiquark type), the contributions
from the gluon condensate are suppressed by large denominators and
would not play any significant roles for the light tetraquark states
\cite{Wang1,Wang2}, the heavy tetraquark state \cite{Wang08072} and
the  heavy molecular state \cite{Wang0904}. In the present case, the
contributions from the gluon condensate $\langle \frac{\alpha_s
GG}{\pi} \rangle $ are rather large,  just like in the sum rules for
the $Y(4660)$ \cite{WangZhang}. If we take a simple replacement
$\bar{s}(x)s(x)\rightarrow \langle \bar{s}s\rangle$ and
$\left[\bar{u}(x)u(x)+\bar{d}(x)d(x) \right]\rightarrow
2\langle\bar{q}q\rangle$ in the interpolating currents $J(x)$ and
$\eta(x)$, we can obtain the standard pseudoscalar heavy quark
current $Q(x)i\gamma_5 Q(x)$, where the gluon condensate $\langle
\frac{\alpha_s GG}{\pi} \rangle $ plays an important rule in the QCD
sum rules \cite{SVZ79}.

In this article, we take the uniform Borel parameter $M^2_{min}$,
i.e. $M^2_{min}\geq 2.7 \, \rm{GeV}^2$ and $M^2_{min}\geq 7.6 \,
\rm{GeV}^2$ in the hidden charm  and hidden bottom channels,
respectively.

In Fig.3, we show the  contributions from the pole terms with
variation of the Borel parameters $M^2$ and the threshold parameters
$s_0$. If the pole dominance criterion is satisfied, the threshold
parameter $s_0$ increases   with  the Borel parameter $M^2$
monotonously. From
 Fig.3-A, we can see that  the pole dominance criterion cannot
be satisfied at the values  $s_0\leq 25 \,\rm{GeV}^2$ and $M^2\geq
2.7\,\rm{GeV}^2$ in the $c\bar{c}s\bar{s}$ channel,  the threshold
parameter $s_0$ has to  be pushed to larger value.

The pole contributions are larger than  $48\%$ at the values $M^2
\leq 3.1 \, \rm{GeV}^2 $ and $s_0\geq
25\,\rm{GeV}^2,\,26\,\rm{GeV}^2$ in the $c\bar{c}q\bar{q}$,
    $c\bar{c}s\bar{s}$
channels respectively; and larger than  $50\%$ at the values $M^2
\leq 8.2 \, \rm{GeV}^2 $, $s_0\geq
142\,\rm{GeV}^2,\,144\,\rm{GeV}^2$ in  the $b\bar{b}q\bar{q}$
  and $b\bar{b}s\bar{s}$ channels respectively. Again we
take the uniform Borel parameter $M^2_{max}$, i.e. $M^2_{max}\leq
3.1 \, \rm{GeV}^2$ and $M^2_{max}\leq 8.2 \, \rm{GeV}^2$ in the
hidden charm  and hidden bottom channels, respectively.

In this article, the threshold parameters are taken as
$s_0=(26\pm1)\,\rm{GeV}^2$, $(27\pm1)\,\rm{GeV}^2$,
$(144\pm2)\,\rm{GeV}^2$ and $(146\pm2)\,\rm{GeV}^2$ in the
$c\bar{c}q\bar{q}$,
    $c\bar{c}s\bar{s}$, $b\bar{b}q\bar{q}$
     and $b\bar{b}s\bar{s}$ channels, respectively;
   the Borel parameters are taken as $M^2=(2.7-3.1)\,\rm{GeV}^2$ and
   $(7.6-8.2)\,\rm{GeV}^2$ in the
hidden charm and hidden bottom channels, respectively. In those
regions, the pole contributions are about  $(48-72)\%$, $(49-72)\%$,
$(50-66)\%$ and $(51-66)\%$ in  the $c\bar{c}s\bar{s}$,
    $c\bar{c}q\bar{q}$, $b\bar{b}s\bar{s}$
     and $b\bar{b}q\bar{q}$ channels, respectively;  the two criteria of the QCD sum rules
are fully  satisfied  \cite{SVZ79,Reinders85}.

The Borel windows $M_{max}^2-M_{min}^2$ change with  variations of
the  threshold parameters $s_0$, see Fig.3. In this article, the
Borel windows  are taken as  $0.4\,\rm{GeV}^2$ and $0.6\,\rm{GeV}^2$
in the hidden charm and hidden bottom channels respectively, they
are small enough. Furthermore, we take uniform Borel windows and
smear the dependence on the threshold parameters $s_0$.  If we take
larger threshold parameters,  the Borel windows are larger and the
resulting  masses are larger, see Fig.4. In this article, we intend
to calculate the possibly  lowest  masses which are supposed to be
the ground state masses  by imposing the two criteria of the QCD sum
rules.

In Fig.4, we plot the  bound state masses $M_Y$ with variation of
the Borel parameters and the threshold parameters.  The hidden charm
current $\bar{c}(x)i\gamma_5 c(x)$ can interpolate the charmonia
$\eta_c$, $\eta_c'$, $\eta_c''$, $\cdots$; and the hidden bottom
current $\bar{b}(x)i\gamma_5 b(x)$ can interpolate the bottomonia
$\eta_b$, $\eta_b'$, $\eta_b''$, $\cdots$ \cite{PDG}. The currents
$J(x)$ have non-vanishing couplings with the bound states $\eta_c
f_0(980)$, $\eta_c'f_0(980)$, $\eta_c''f_0(980)$, $\cdots$ and
$\eta_b f_0(980)$, $\eta_b'f_0(980)$, $\eta_b''f_0(980)$,
$\eta_b'''f_0(980)$, $\cdots$, respectively. The mass of the
$\eta_b$ listed in the Particle Data Group is
$M_{\eta_b}=9388.9^{+3.1}_{-2.3}\pm2.7\,\rm{MeV}$, while the
$\eta_b'$, $\eta_b''$, $\eta_b'''$, $\cdots$ are not observed yet
\cite{PDG}. In the constituent quark models, the mass splitting
between the spin-singlet and spin-triplet are proportional to
$\frac{\sigma_i \cdot\sigma_j}{M_i M_j}$, in the heavy quark limit,
the $\eta_b$ and $\Upsilon$ degenerate. The constituent quark mass
$M_b$ is large enough, $M_\Upsilon=(9460.30\pm0.26)\,\rm{MeV}$, the
energy gap between the $\eta_b$ and $\Upsilon$ is about
$71.4\,\rm{MeV}$, the energy gaps between the radial excited states
are even smaller.  In this article, we assume the masses of the
$\eta_b$, $\eta_b'$, $\eta_b''$, $\cdots$ are slightly smaller than
ones of the $\Upsilon$, $\Upsilon'$, $\Upsilon''$, $\cdots$
respectively.

From Figs.3-A,3-C,4-A,4-C, we can see that the QCD sum rules support
existence of the $\eta_c' f_0(980)$ and $\eta_b'''f_0(980)$ bound
states, the nominal thresholds  of the $\eta_c-f_0(980)$ and
$\Upsilon''-f_0(980)$ systems are too low, and we cannot reproduce
the $\eta_c f_0(980)$ and $\eta_b''f_0(980)$ bound states. Our
 predictions for the masses the Y(4660) \cite{WangZhang} and  $Y(4616)$ support the conjecture of Voloshin et
al, i.e. a formation of hadro-charmonium is favored for higher
charmonium resonances $\psi'$ and $\chi_{cJ}$ as compared to the
lowest states $J/\psi$ and $\eta_c$ \cite{Voloshin0803}.

In this article, we  intend to prove that the $\eta_c' f_0(980)$
bound state can be reproduced by the QCD sum rules, the pseudoscalar
charmonium-like state $Y(4616)$ predicted in Ref.\cite{GuoPRL} maybe
exist.

Taking into account all uncertainties of the input parameters,
finally we obtain the values of the masses and pole resides of
 the  pseudoscalar  bound states  $Y$, which are  shown in Figs.5-6 and Tables 1-2.
In this article,  we calculate the uncertainties $\delta$  with the
formula
\begin{eqnarray}
\delta=\sqrt{\sum_i\left(\frac{\partial f}{\partial
x_i}\right)^2\mid_{x_i=\bar{x}_i} (x_i-\bar{x}_i)^2}\,  ,
\end{eqnarray}
 where the $f$ denote  the
hadron mass  $M_Y$ and the pole residue $\lambda_Y$,  the $x_i$
denote the input QCD parameters $m_c$, $m_b$, $\langle \bar{q}q
\rangle$, $\langle \bar{s}s \rangle$, $\cdots$, and the threshold
parameter $s_0$ and Borel parameter $M^2$. As the partial
 derivatives   $\frac{\partial f}{\partial x_i}$ are difficult to carry
out analytically, we take the  approximation $\left(\frac{\partial
f}{\partial x_i}\right)^2 (x_i-\bar{x}_i)^2\approx
\left[f(\bar{x}_i\pm \Delta x_i)-f(\bar{x}_i)\right]^2$ in the
numerical calculations.

\begin{table}
\begin{center}
\begin{tabular}{|c|c|c|c|}
\hline\hline bound states & $M_Y$ ($ \rm{GeV}$) & $M_{\eta_c'/\Upsilon'''}+M_{f_0/\sigma}$ ($ \rm{GeV}$)\\
\hline
      $c\bar{c}s\bar{s}$  &$4.68\pm0.29$ &$4.617$\\ \hline
           $c\bar{c}q\bar{q} $ &$4.56\pm0.21$& $4.037-4.837$\\      \hline
    $b\bar{b}s\bar{s}$  &$11.42\pm0.21$ &$11.559$\\ \hline
            $ b\bar{b}q\bar{q} $ &$11.36\pm0.18$ &$10.979-11.779$\\      \hline
    \hline
\end{tabular}
\end{center}
\caption{ The masses  of the pseudoscalar bound  states. }
\end{table}

\begin{table}
\begin{center}
\begin{tabular}{|c|c|c|}
\hline\hline bound states & $\lambda_{Y}$ ($10^{-2} \rm{GeV}^5$)\\
\hline
      $c\bar{c}s\bar{s}$  &$3.63\pm1.80$\\ \hline
            $c\bar{c}q\bar{q} $ & $3.41\pm1.37$\\      \hline
    $b\bar{b}s\bar{s}$   &$20.7\pm8.0$\\ \hline
            $ b\bar{b}q\bar{q} $  &$20.4\pm6.5$\\      \hline
    \hline
\end{tabular}
\end{center}
\caption{ The  pole residues of the pseudoscalar bound  states. }
\end{table}

In table 1, we also present the  nominal thresholds of the
$\eta_c'-f_0(980)$, $\eta_c'-\sigma(400-1200)$,
$\Upsilon'''-f_0(980)$ and $\Upsilon'''-\sigma(400-1200)$ systems.
From the table, we can see that  there maybe exist a bound state
$\eta_c'f_0(980)$ as the  partner of the $Y(4660)$. The predicted
mass of the  bound state $\eta_c'\sigma(400-1200)$ is about
$(4.56\pm0.21)\rm{GeV}$, while the nominal threshold of the
$\eta_c'-\sigma(400-1200)$ system is about $(4.037-4.837)\,
\rm{GeV}$. There maybe exist such a bound state. The
 bound states $\eta_c'f_0(980)$ and $\eta_c'\sigma(400-1200)$ can  be produced  in the
exclusive decays of the $B$ meson through $b\rightarrow
c\bar{c}s,\,c\bar{c}q$ at the quark level.

In the $b\bar{b}s\bar{s}$ channel, the numerical result
$M_Y=11.42\pm0.21\,\rm{GeV}$ indicates that there maybe exist a
 bound state $\Upsilon'''f_0(980)$, which is consistent with the
nominal threshold $M_{\Upsilon'''}+M_{f_0}=11.559\,\rm{GeV}$, while
the  nominal thresholds $M_{\Upsilon}+M_{f_0}=10.44\,\rm{GeV}$,
$M_{\Upsilon'}+M_{f_0}=11.00\,\rm{GeV}$,
$M_{\Upsilon''}+M_{f_0}=11.335\,\rm{GeV}$ are too low. The scalar
meson $\sigma(400-1200)$ is rather broad with the Breit-Wigner mass
formula $(400-1200)-i(250-500)$ \cite{PDG}. Considering the $SU(3)$
symmetry of the light flavor quarks, we can obtain the conclusion
tentatively that there maybe exist the  bound states
$\eta_c'\sigma(400-1200)$ and $\eta_c'''\sigma(400-1200)$ which lie
in the regions $(4.037-4.837) \,\rm{GeV}$ and
$(10.979-11.779)\,\rm{GeV}$, respectively. As the energy gaps
between the $\Upsilon$'s are rather small and the scalar meson
$\sigma(400-1200)$ is broad enough, there maybe   exist the
$\eta_b\sigma(400-1200)$, $\eta_b'\sigma(400-1200)$ and
$\eta_b''\sigma(400-1200)$ bound states. We cannot draw decisive
conclusion with the QCD sum rules alone.

The LHCb is a dedicated $b$ and $c$-physics precision experiment at
the LHC (large hadron collider). The LHC will be the world's most
copious  source of the $b$ hadrons, and  a complete spectrum of the
$b$ hadrons will be available through gluon fusion. In proton-proton
collisions at $\sqrt{s}=14\,\rm{TeV}$, the $b\bar{b}$ cross section
is expected to be $\sim 500\mu b$ producing $10^{12}$ $b\bar{b}$
pairs in a standard  year of running at the LHCb operational
luminosity of $2\times10^{32} \rm{cm}^{-2} \rm{sec}^{-1}$
\cite{LHC}. The pseudoscalar  bound  states $\eta_b'''f_0(980)$ and
$\eta_b'''\sigma(400-1200)$ predicted in the present work may be
observed at the LHCb, if they exist indeed. We can search for those
bound states in the $\eta_b\pi\pi$, $\eta_b'\pi\pi$,
$\eta_b''\pi\pi$, $\eta_b'''\pi\pi$, $\eta_b K\bar{K}$,
$\eta_b'K\bar{K}$, $\eta_b''K\bar{K}$, $\eta_b'''K\bar{K}$, $\cdots$
invariant mass distributions.

\section{Conclusion}
In this article, we study the pseudoscalar  bound state
$\eta_c'f_0(980)$ (irrespective of the hadro-charmonium and the
molecular state)
 with  the QCD sum rules, the numerical result
$M_Y=4.68\pm0.29\,\rm{GeV}$ is consistent with the value
$4616^{+5}_{-6}\,\rm{MeV}$ predicted by Guo et al. Considering the
$SU(3)$ symmetry of the light flavor quarks and the heavy quark
symmetry, we also study the bound states $\eta_c'\sigma(400-1200)$,
$\eta_b'''f_0(980)$ and $\eta_b'''\sigma(400-1200)$ with the QCD sum
rules, and make reasonable  predictions for their masses. Our
predictions depend heavily on  the two criteria (pole dominance and
convergence of the operator product expansion) of the QCD sum rules.
We can search for those bound states at the LHCb, the KEK-B or the
Fermi-lab Tevatron.

\begin{figure}
 \centering
 \includegraphics[totalheight=4.5cm,width=6cm]{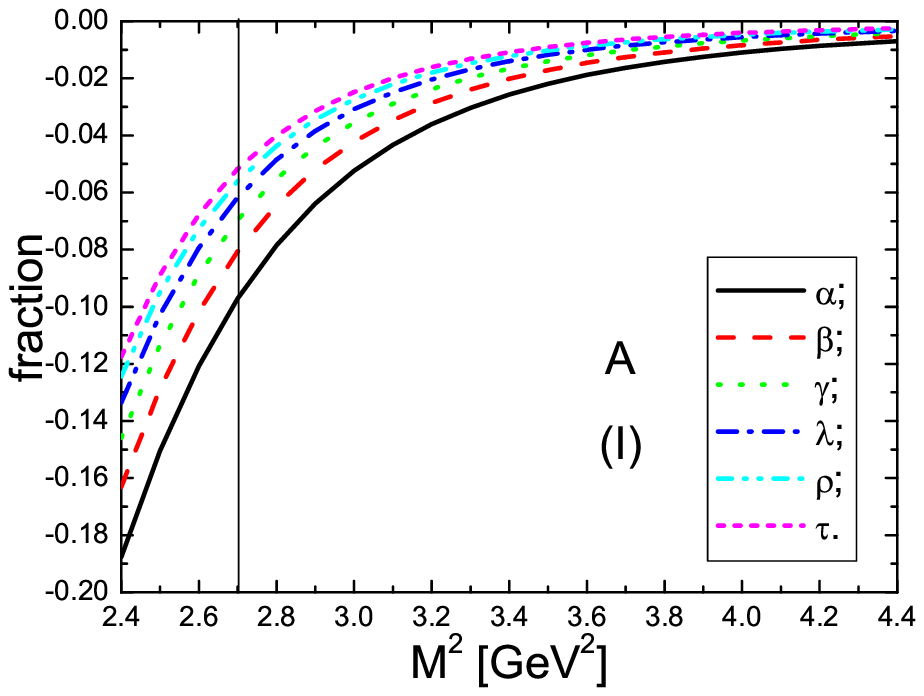}
 \includegraphics[totalheight=4.5cm,width=6cm]{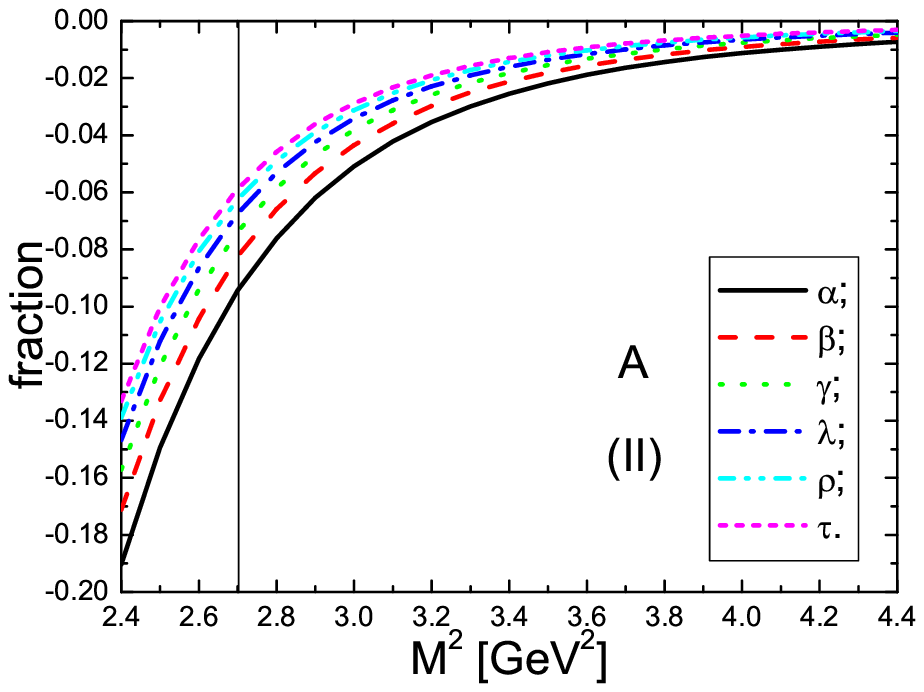}
 \includegraphics[totalheight=4.5cm,width=6cm]{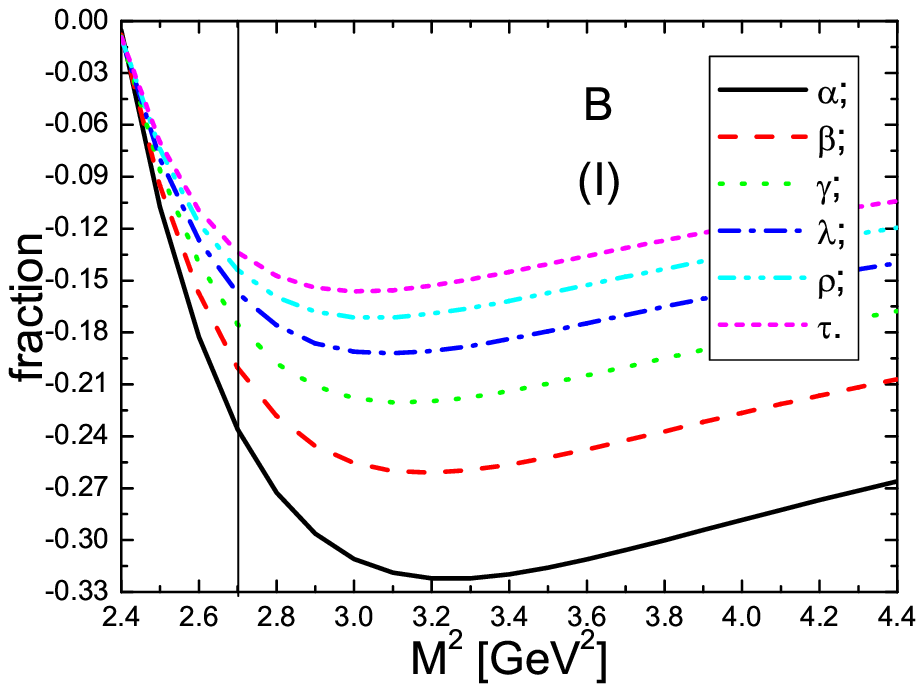}
 \includegraphics[totalheight=4.5cm,width=6cm]{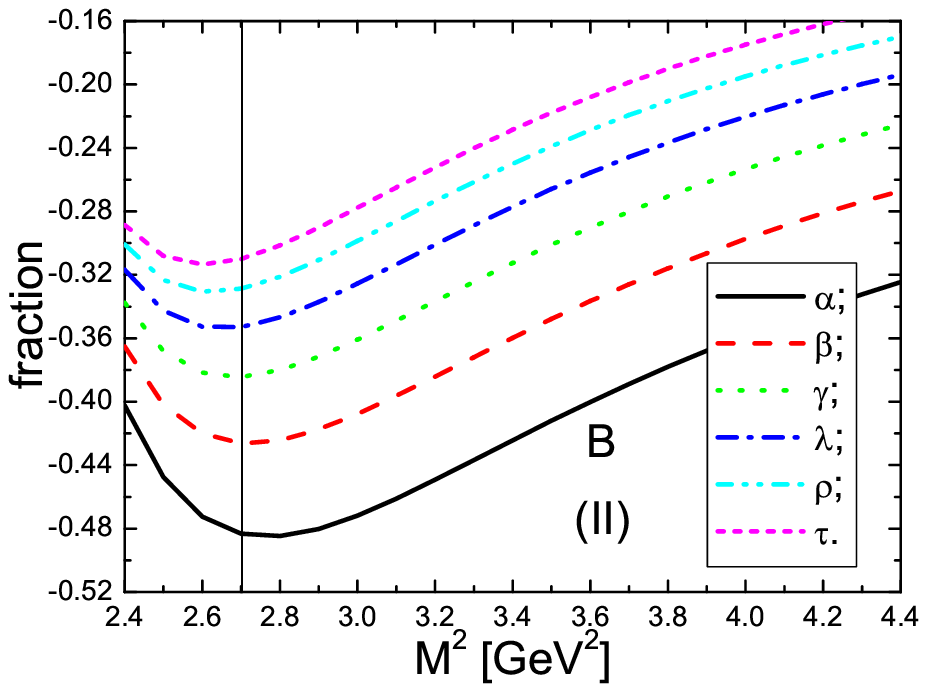}
 \includegraphics[totalheight=4.5cm,width=6cm]{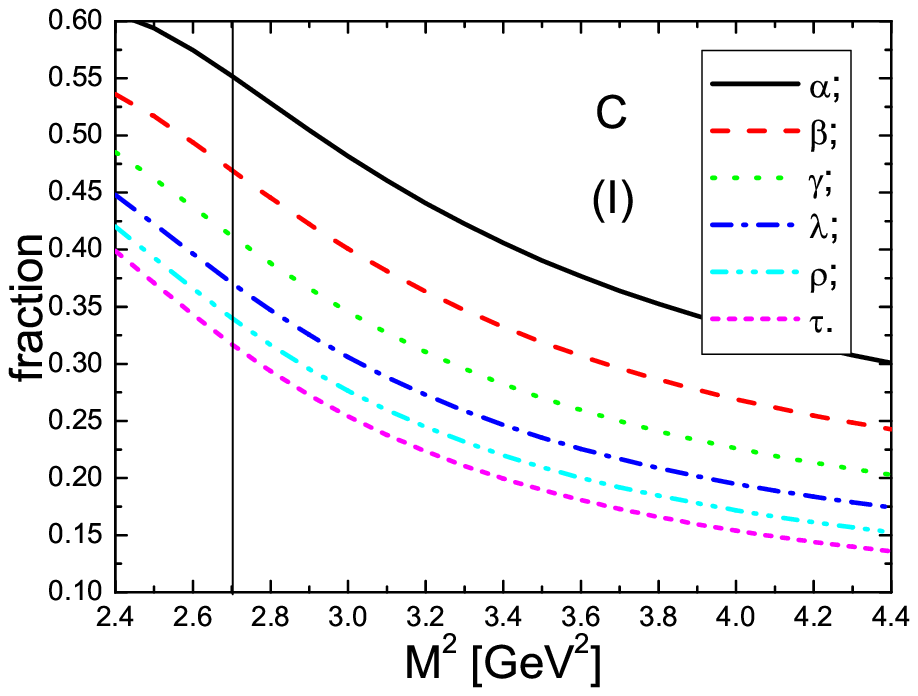}
 \includegraphics[totalheight=4.5cm,width=6cm]{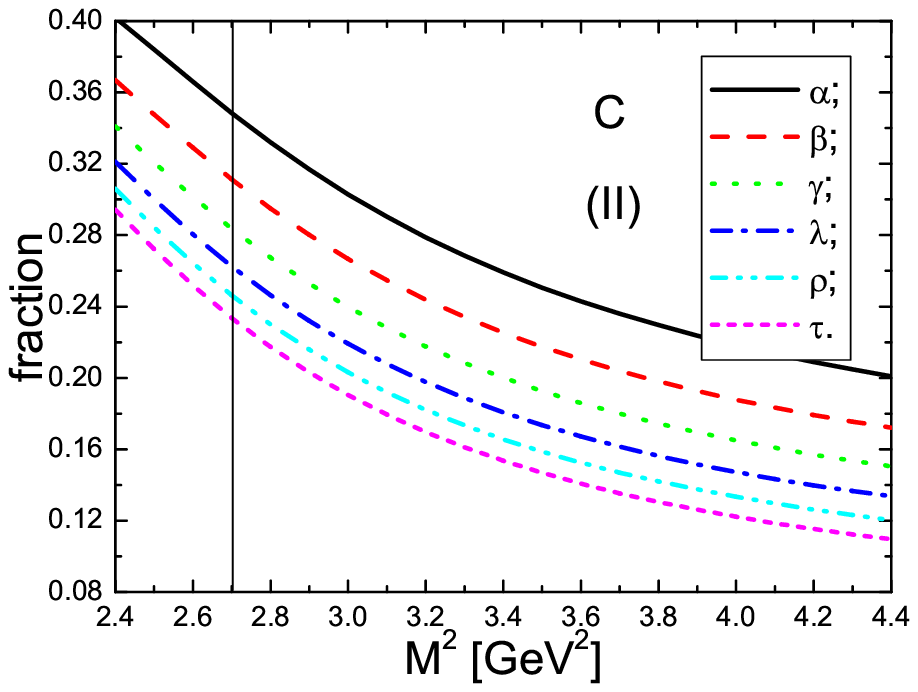}
     \caption{ The contributions from different terms with variation of the Borel
   parameter $M^2$  in the operator product expansion. The $A$,
   $B$ and $C$   correspond to the contributions from the $\langle \bar{q}g_s\sigma G q \rangle^2$ term,
   the $\langle \bar{q} q \rangle^2$ +$\langle \bar{q} q \rangle\langle \bar{q}g_s\sigma G q \rangle$
   term and  the $\langle \frac{\alpha_s GG}{\pi} \rangle $ term, respectively. The
   (I) and (II)  denote the $c\bar{c}s\bar{s}$
  and $c\bar{c}q\bar{q}$ channels, respectively. The
notations
   $\alpha$, $\beta$, $\gamma$, $\lambda$, $\rho$ and $\tau$  correspond to the threshold
   parameters $s_0=23\,\rm{GeV}^2$,
   $24\,\rm{GeV}^2$, $25\,\rm{GeV}^2$, $26\,\rm{GeV}^2$, $27\,\rm{GeV}^2$ and $28\,\rm{GeV}^2$, respectively.}
\end{figure}

\begin{figure}
 \centering
 \includegraphics[totalheight=4.5cm,width=6cm]{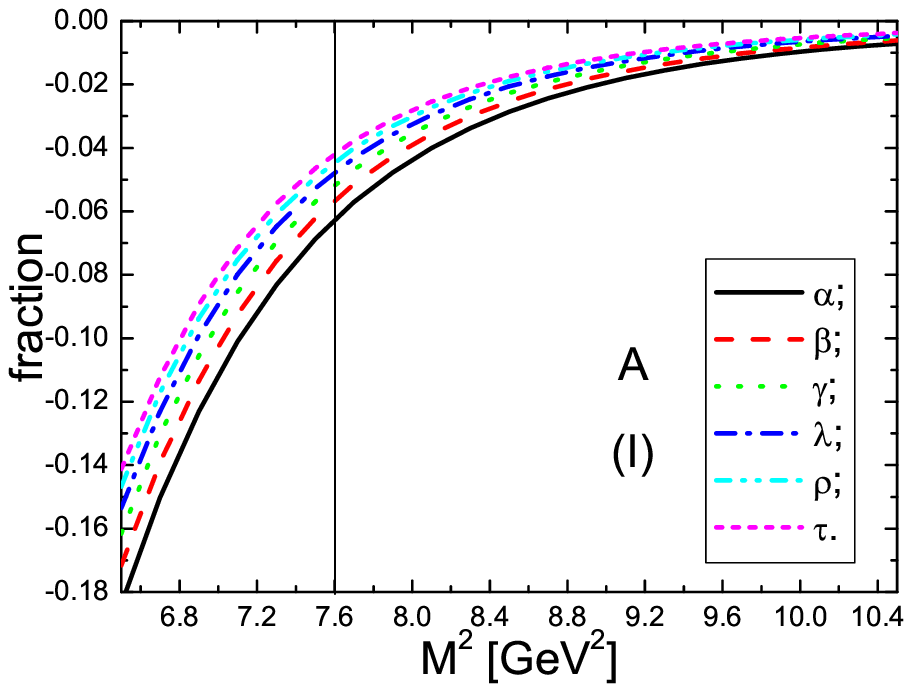}
 \includegraphics[totalheight=4.5cm,width=6cm]{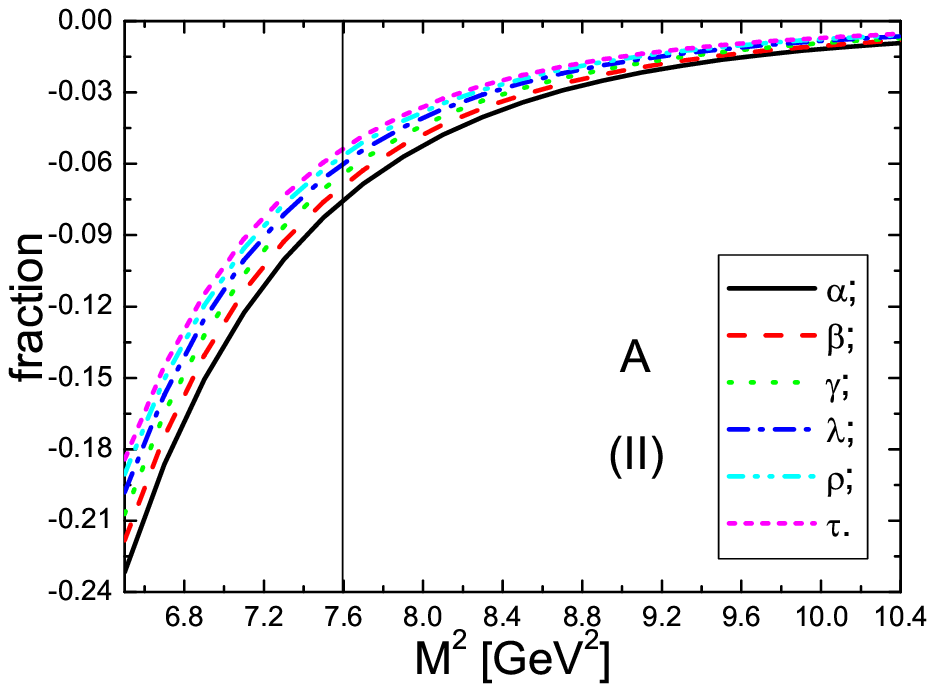}
 \includegraphics[totalheight=4.5cm,width=6cm]{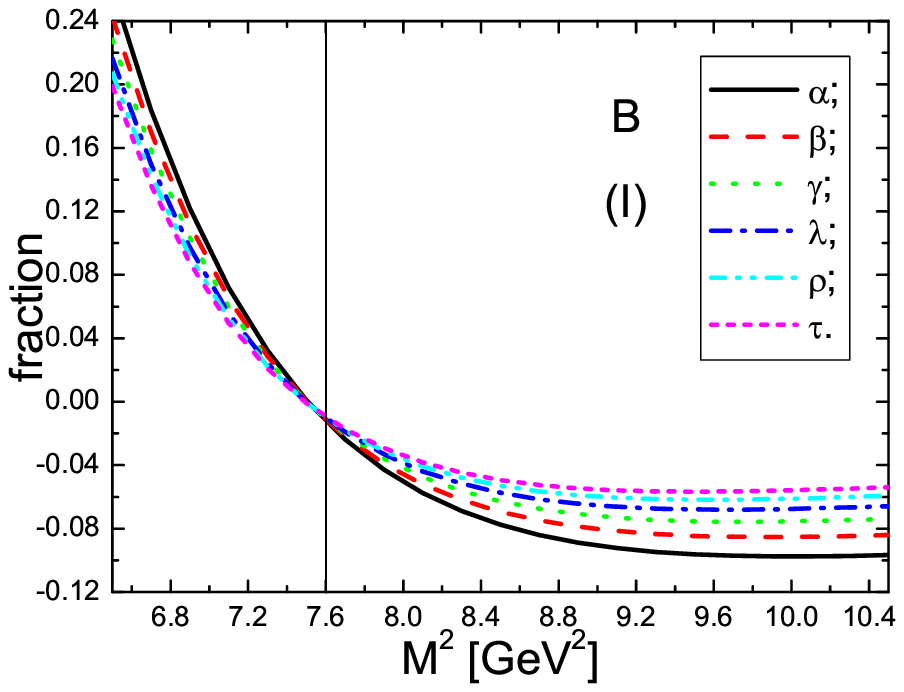}
 \includegraphics[totalheight=4.5cm,width=6cm]{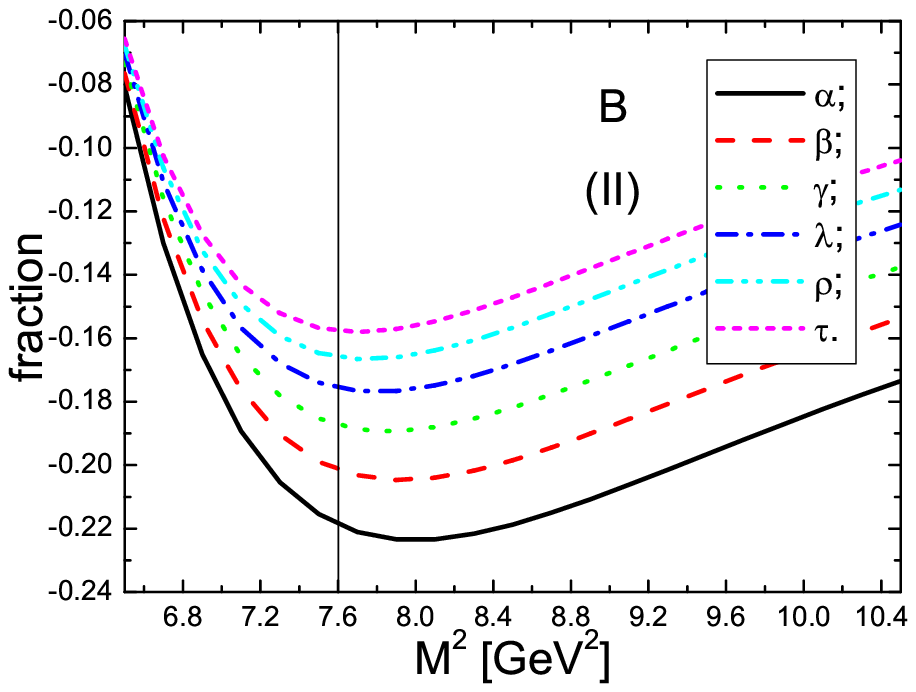}
 \includegraphics[totalheight=4.5cm,width=6cm]{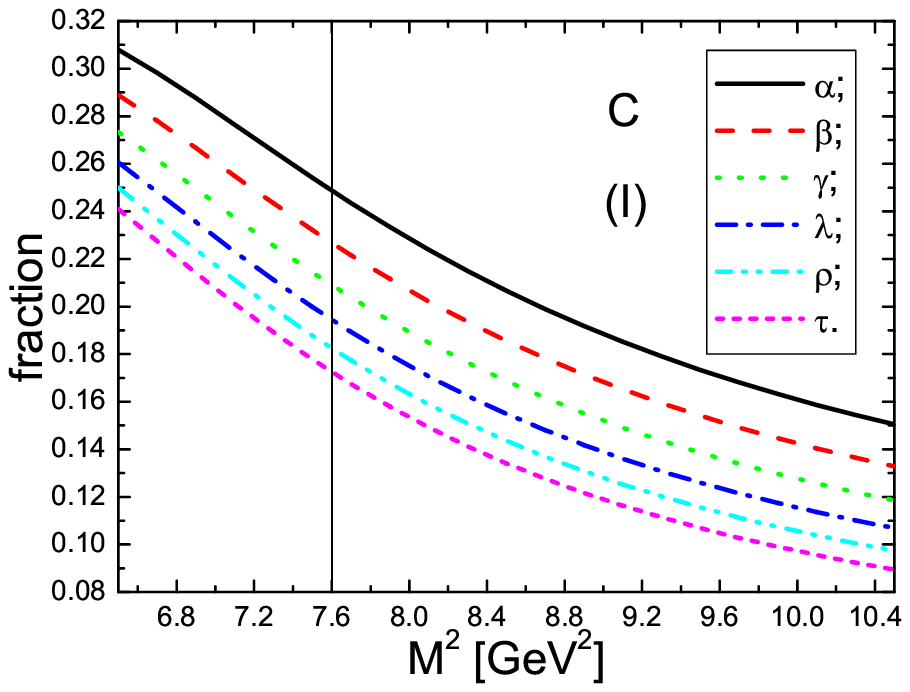}
 \includegraphics[totalheight=4.5cm,width=6cm]{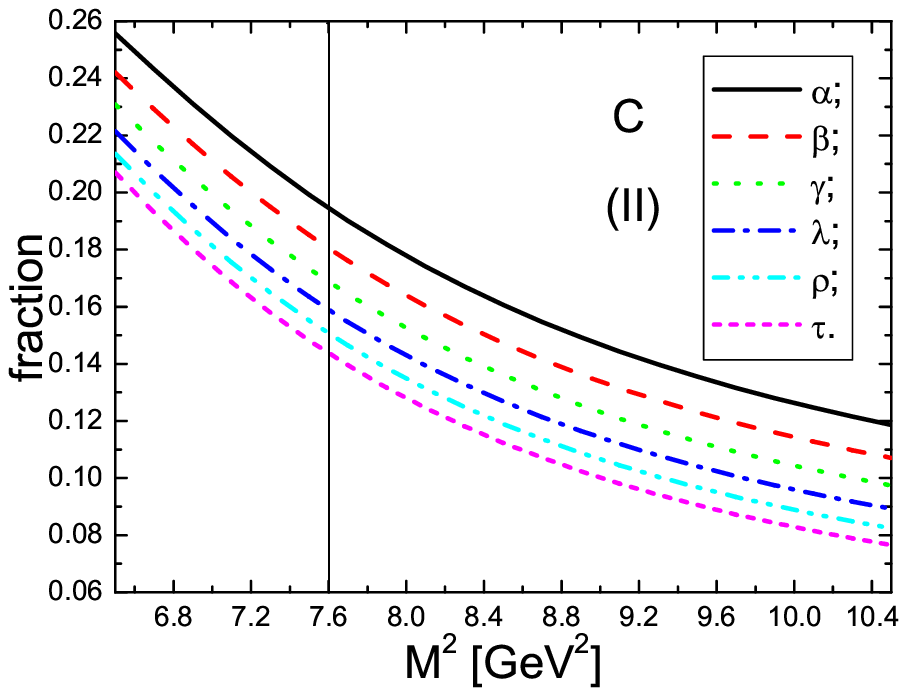}
     \caption{ The contributions from different terms with variation of the Borel
   parameter $M^2$  in the operator product expansion. The $A$,
   $B$ and $C$   correspond to the contributions from the $\langle \bar{q}g_s\sigma G q \rangle^2$ term,
   the $\langle \bar{q} q \rangle^2$ +$\langle \bar{q} q \rangle\langle \bar{q}g_s\sigma G q \rangle$ term
   and the $\langle \frac{\alpha_s GG}{\pi} \rangle $ term, respectively. The
   (I) and (II)  denote the $b\bar{b}s\bar{s}$
  and $b\bar{b}q\bar{q}$ channels, respectively. The
notations
   $\alpha$, $\beta$, $\gamma$, $\lambda$, $\rho$ and $\tau$  correspond to the threshold
   parameters $s_0=138\,\rm{GeV}^2$,
   $140\,\rm{GeV}^2$, $142\,\rm{GeV}^2$, $144\,\rm{GeV}^2$, $146\,\rm{GeV}^2$ and $148\,\rm{GeV}^2$, respectively.}
\end{figure}

\begin{figure}
 \centering
 \includegraphics[totalheight=5cm,width=6cm]{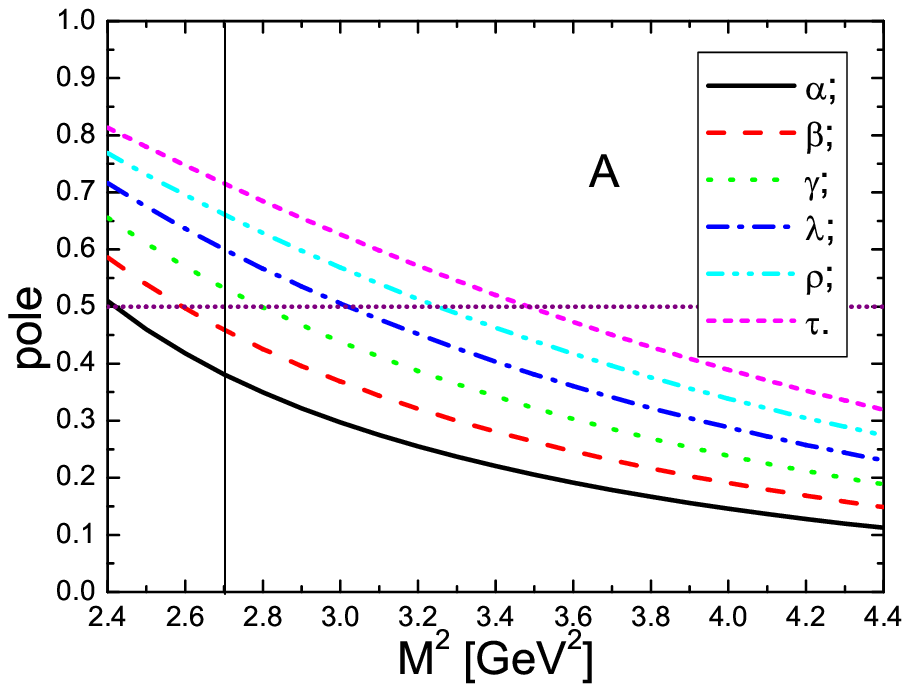}
\includegraphics[totalheight=5cm,width=6cm]{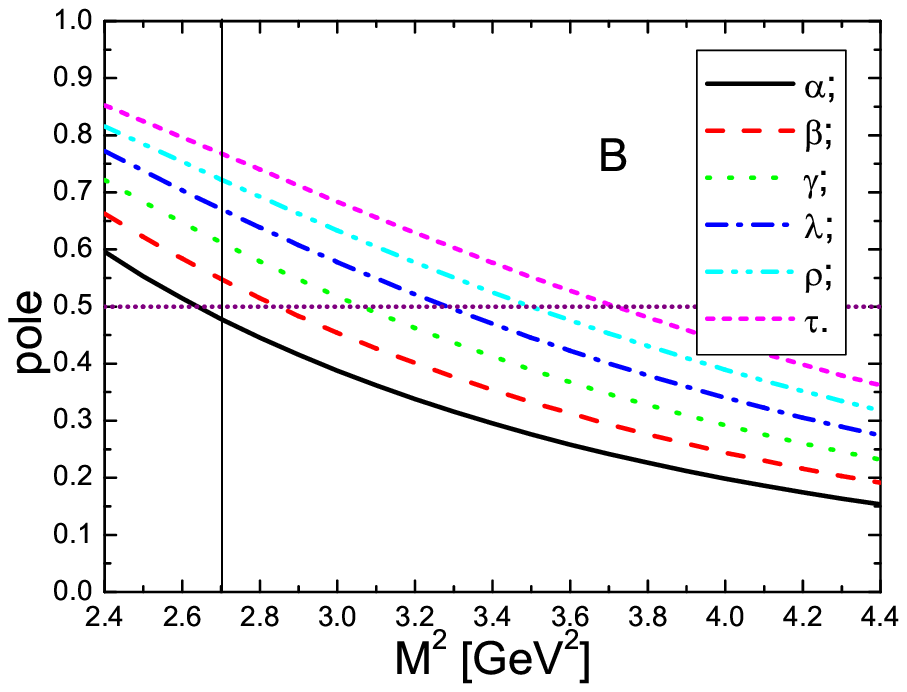}
\includegraphics[totalheight=5cm,width=6cm]{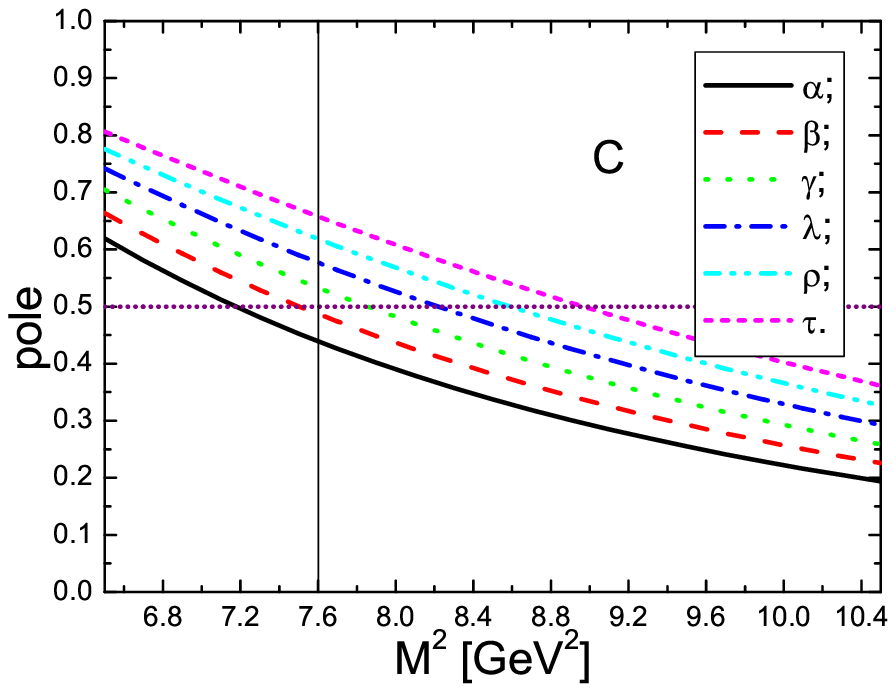}
\includegraphics[totalheight=5cm,width=6cm]{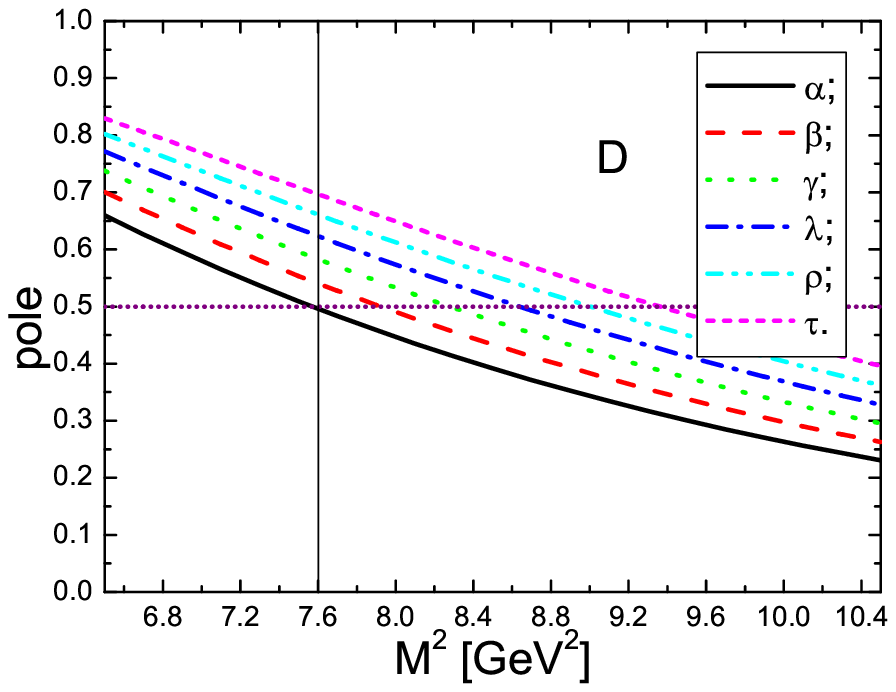}
   \caption{ The contributions from the pole terms with variation of the Borel parameter $M^2$. The $A$, $B$, $C$,
   and $D$ denote the $c\bar{c}s\bar{s}$,
    $c\bar{c}q\bar{q}$, $b\bar{b}s\bar{s}$
    and $b\bar{b}q\bar{q}$ channels, respectively.  In the hidden charm channels, the notations
   $\alpha$, $\beta$, $\gamma$, $\lambda$, $\rho$ and $\tau$  correspond to the threshold
   parameters $s_0=23\,\rm{GeV}^2$,
   $24\,\rm{GeV}^2$, $25\,\rm{GeV}^2$, $26\,\rm{GeV}^2$, $27\,\rm{GeV}^2$ and $28\,\rm{GeV}^2$, respectively
   ;  while in the hidden bottom channels they correspond to
    the threshold
   parameters  $s_0=138\,\rm{GeV}^2$,
   $140\,\rm{GeV}^2$, $142\,\rm{GeV}^2$, $144\,\rm{GeV}^2$, $146\,\rm{GeV}^2$ and $148\,\rm{GeV}^2$, respectively.}
\end{figure}

\begin{figure}
 \centering
  \includegraphics[totalheight=5cm,width=6cm]{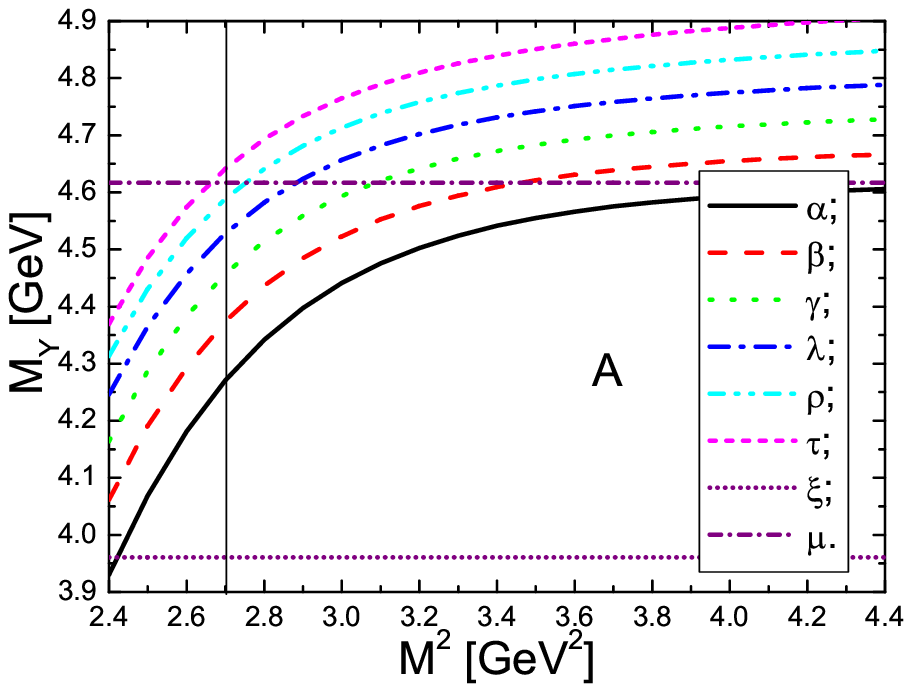}
  \includegraphics[totalheight=5cm,width=6cm]{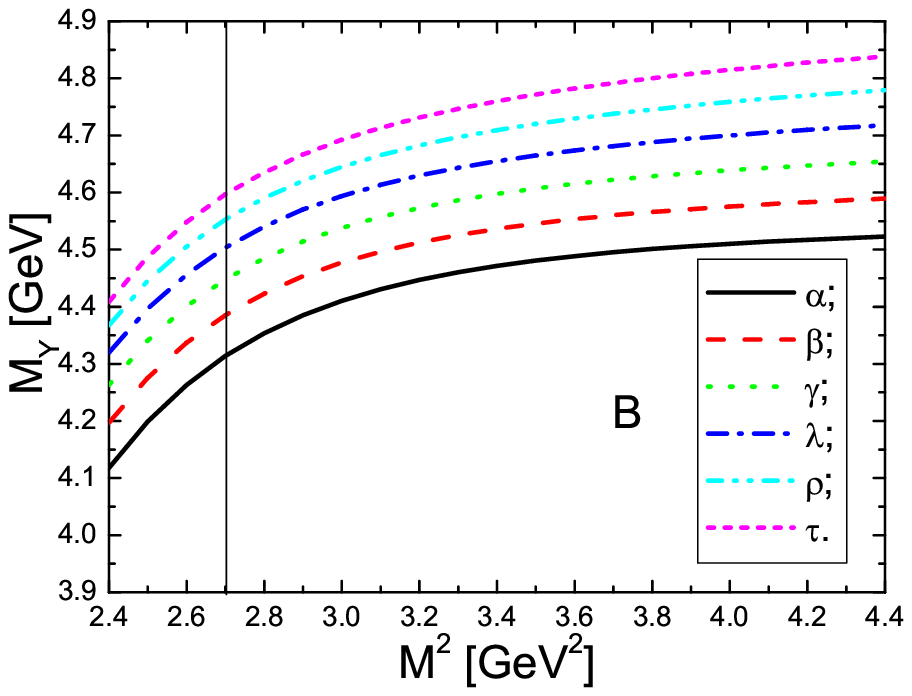}
  \includegraphics[totalheight=5cm,width=6cm]{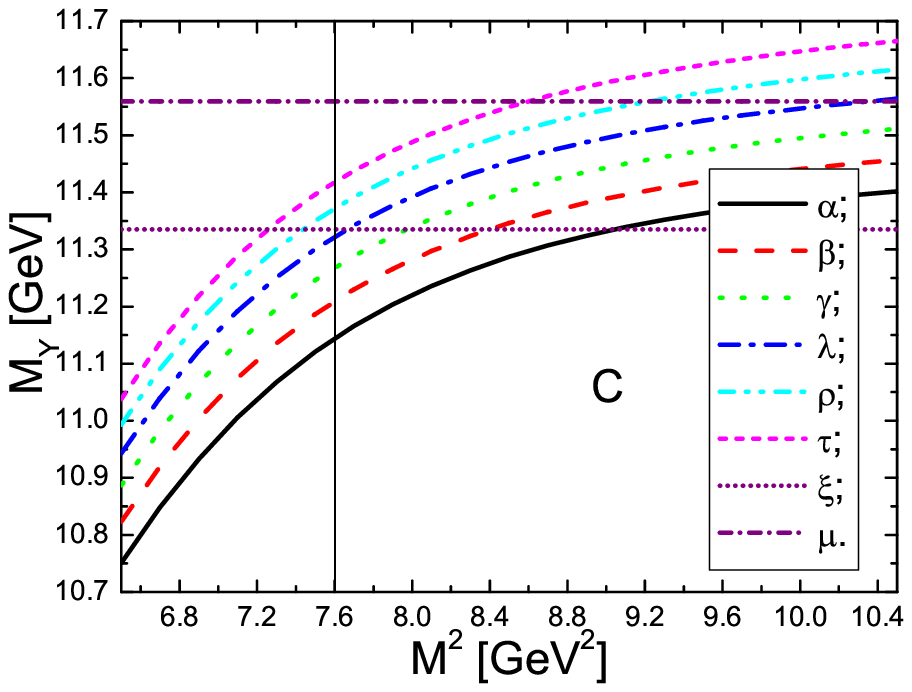}
  \includegraphics[totalheight=5cm,width=6cm]{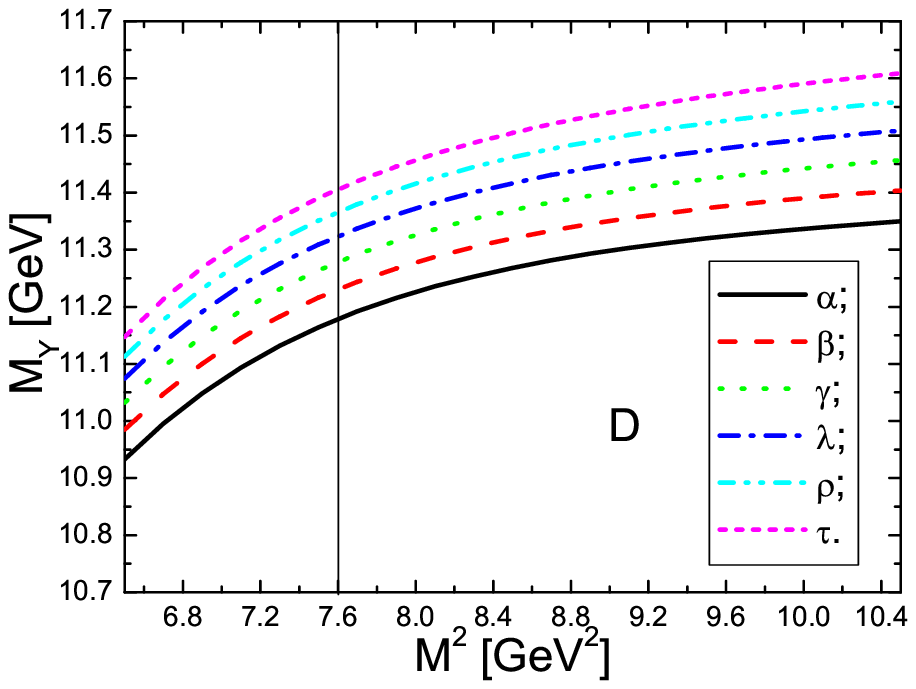}
   \caption{ The masses of the pseudoscalar  bound  states with variation of the Borel parameter $M^2$ and threshold parameter
   $s_0$. The $A$, $B$, $C$,
   and $D$ denote the $c\bar{c}s\bar{s}$,
    $c\bar{c}q\bar{q}$, $b\bar{b}s\bar{s}$,
   and $b\bar{b}q\bar{q}$ channels, respectively. In the hidden charm channels, the notations
   $\alpha$, $\beta$, $\gamma$, $\lambda$, $\rho$ and $\tau$  correspond to the threshold
   parameters $s_0=23\,\rm{GeV}^2$,
   $24\,\rm{GeV}^2$, $25\,\rm{GeV}^2$, $26\,\rm{GeV}^2$, $27\,\rm{GeV}^2$ and $28\,\rm{GeV}^2$, respectively
   ;  while in the hidden bottom channels they correspond to
    the threshold
   parameters  $s_0=138\,\rm{GeV}^2$,
   $140\,\rm{GeV}^2$, $142\,\rm{GeV}^2$, $144\,\rm{GeV}^2$, $146\,\rm{GeV}^2$ and $148\,\rm{GeV}^2$, respectively. The $\xi$ and $\mu$
   denote the $\eta_c -f_0(980)$ and $\eta_c'-f_0(980)$ thresholds respectively in the $c\bar{c}s\bar{s}$ channel,
   while in the $b\bar{b}s\bar{s}$ channel they
   correspond to $\Upsilon''- f_0(980)$ and $\Upsilon'''- f_0(980)$ thresholds respectively.}
\end{figure}

\begin{figure}
 \centering
  \includegraphics[totalheight=5cm,width=6cm]{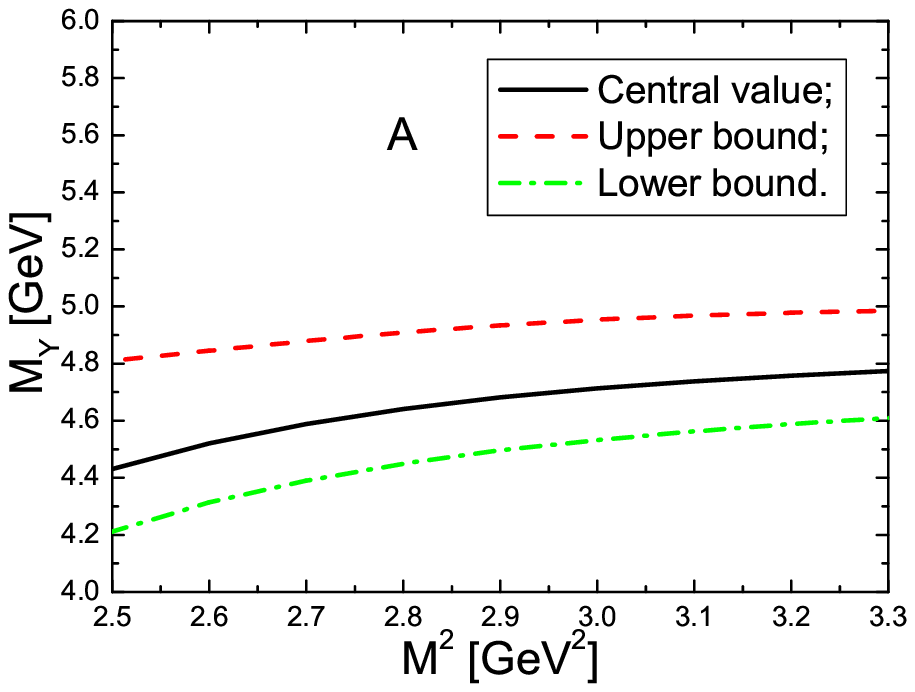}
  \includegraphics[totalheight=5cm,width=6cm]{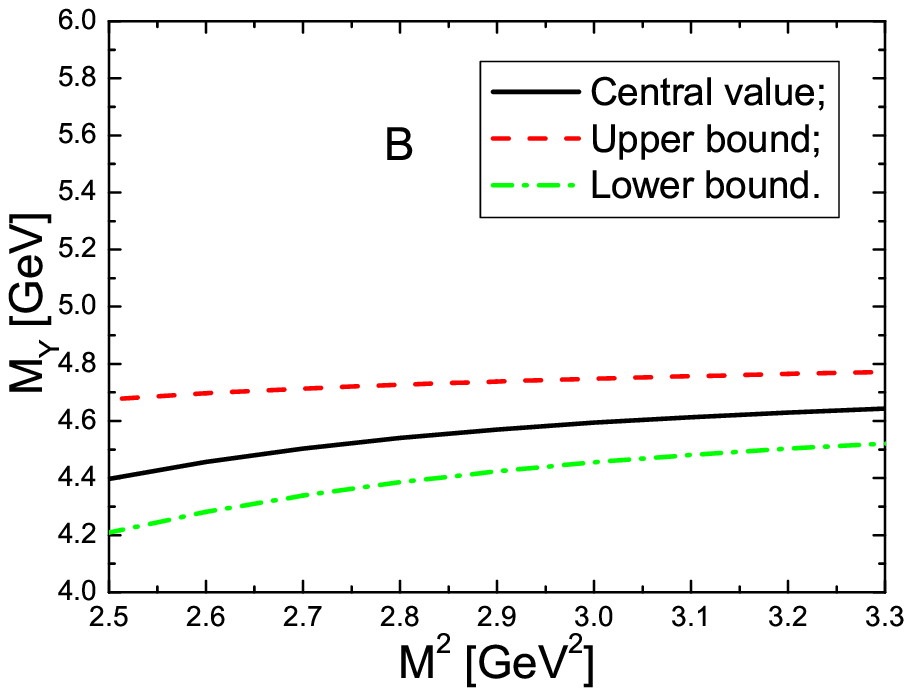}
  \includegraphics[totalheight=5cm,width=6cm]{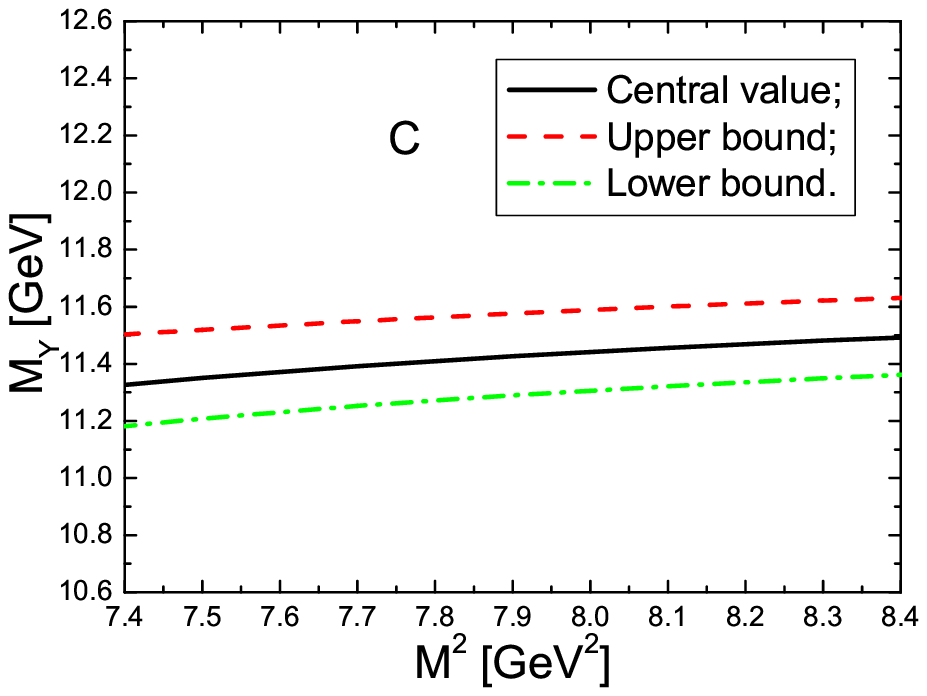}
  \includegraphics[totalheight=5cm,width=6cm]{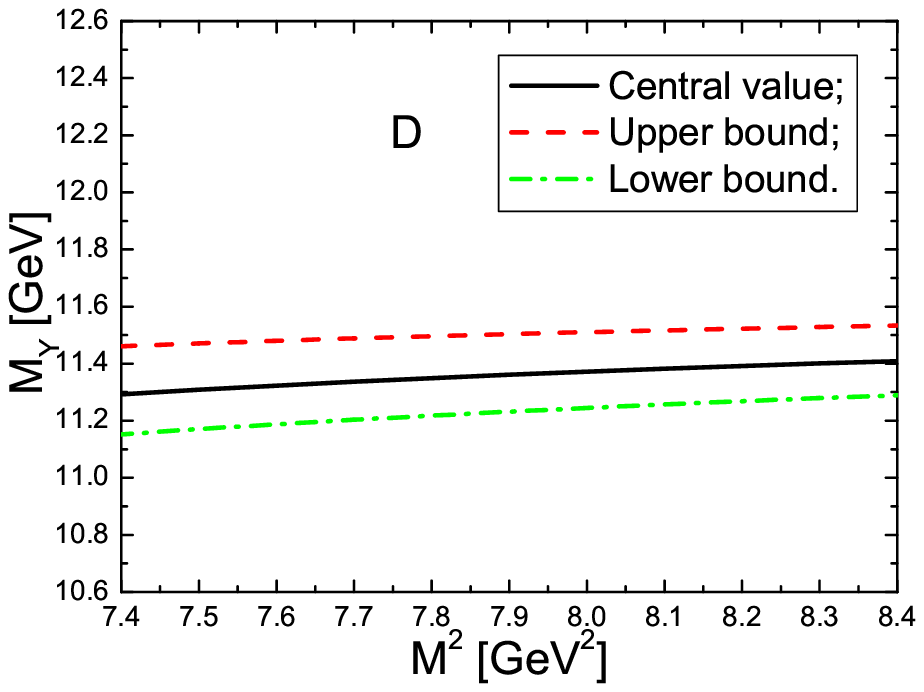}
   \caption{ The masses of the pseudoscalar bound  states with variation of the Borel parameter $M^2$. The $A$, $B$, $C$,
   and $D$ denote the $c\bar{c}s\bar{s}$,
    $c\bar{c}q\bar{q}$, $b\bar{b}s\bar{s}$,
   and $b\bar{b}q\bar{q}$ channels, respectively. }
\end{figure}

\begin{figure}
 \centering
  \includegraphics[totalheight=5cm,width=6cm]{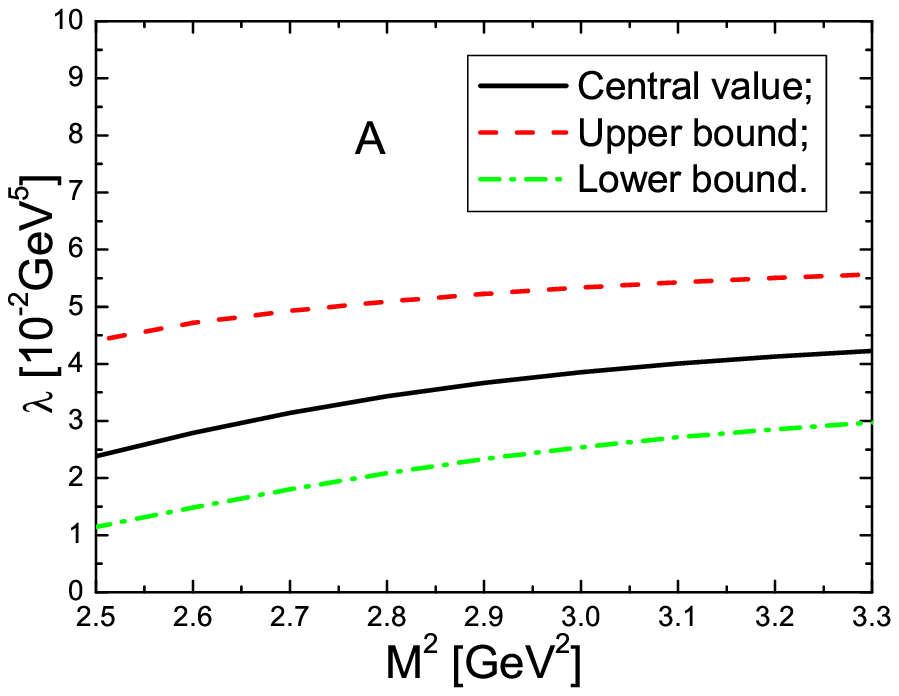}
    \includegraphics[totalheight=5cm,width=6cm]{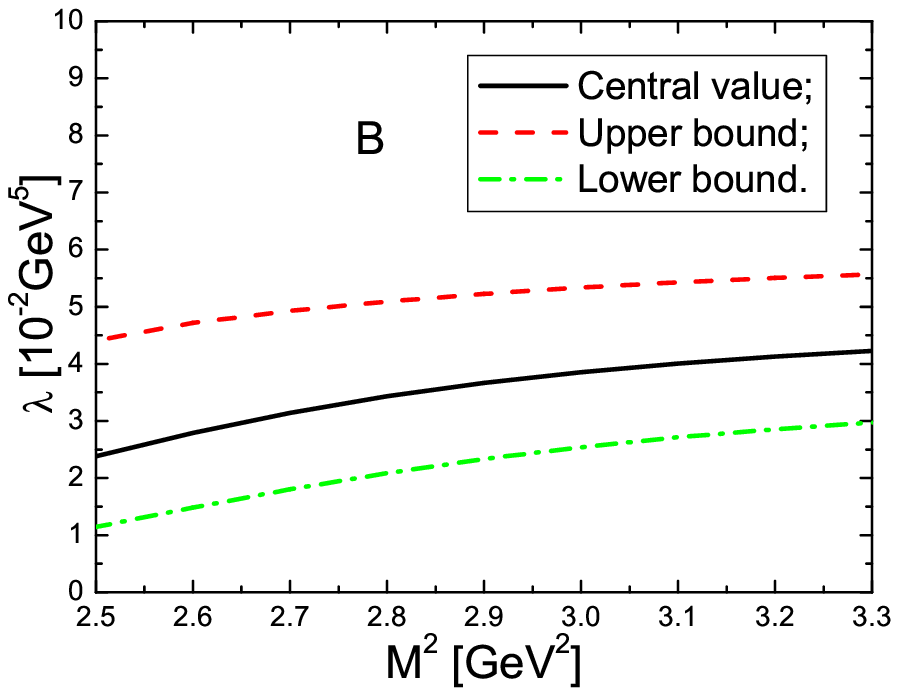}
    \includegraphics[totalheight=5cm,width=6cm]{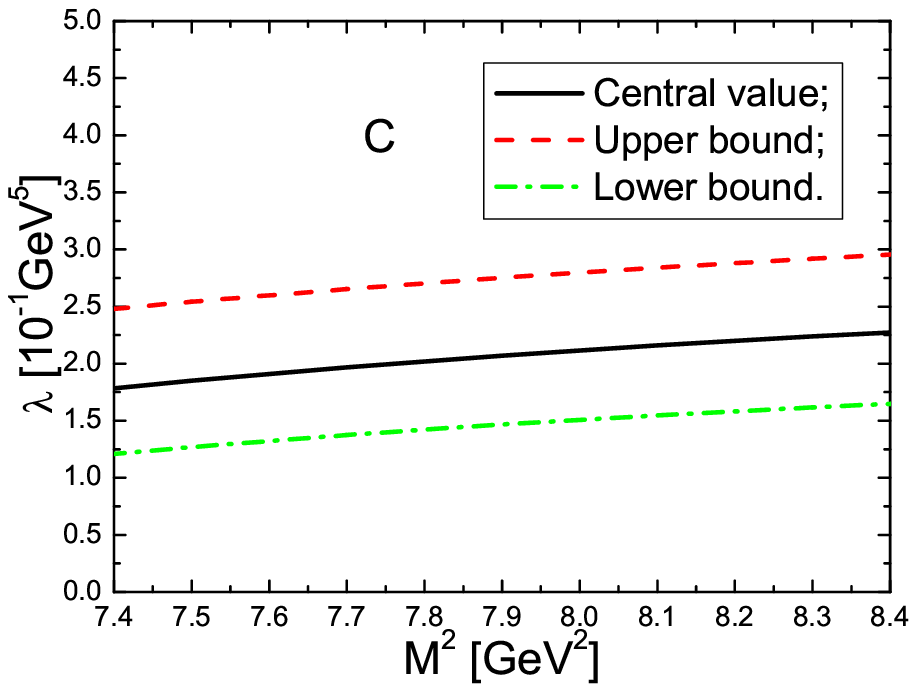}
    \includegraphics[totalheight=5cm,width=6cm]{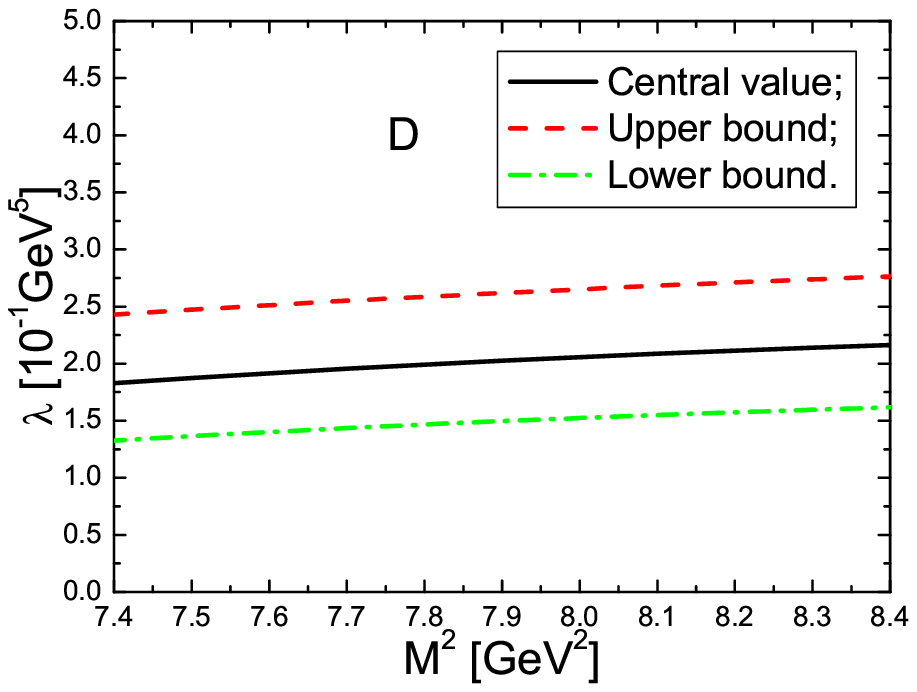}
     \caption{ The pole residues of the pseudoscalar  bound states with variation of the Borel parameter $M^2$. The $A$, $B$, $C$,
   and $D$ denote the $c\bar{c}s\bar{s}$,
    $c\bar{c}q\bar{q}$, $b\bar{b}s\bar{s}$,
   and $b\bar{b}q\bar{q}$ channels, respectively.}
\end{figure}

\section*{Appendix}
The spectral densities at the level of the quark-gluon degrees of
freedom:
\begin{eqnarray}
 \rho_0(s)&=&\frac{3}{2048 \pi^6}
\int_{\alpha_{i}}^{\alpha_{f}}d\alpha \int_{\beta_{i}}^{1-\alpha}
d\beta\alpha\beta(1-\alpha-\beta)^2(s-\widetilde{m}^2_Q)^3(3s+\widetilde{m}^2_Q)\,,
\end{eqnarray}

\begin{eqnarray}
\rho_{\langle
\bar{s}s\rangle}(s)&=&\frac{9m_s\langle\bar{s}s\rangle}{32\pi^4}
\int_{\alpha_{i}}^{\alpha_{f}}d\alpha \int_{\beta_{i}}^{1-\alpha}
d\beta\alpha\beta s(s-\widetilde{m}^2_Q)\nonumber \\
&&-\frac{m_s\langle\bar{s}g_s \sigma Gs\rangle}{32 \pi^4}
\int_{\alpha_{i}}^{\alpha_{f}}d\alpha \alpha(1-\alpha)
(3s-\widetilde{\widetilde{m}}^2_Q) \,,
\end{eqnarray}

\begin{eqnarray}
\rho_{\langle \bar{s}s\rangle^2}(s)&=&-\frac{ \langle\bar{s}
s\rangle^2}{16 \pi^2} \int_{\alpha_{i}}^{\alpha_{f}}d\alpha
\alpha(1-\alpha)  (3s-\widetilde{\widetilde{m}}^2_Q)  \nonumber \\
&& +\frac{ \langle\bar{s} s\rangle \langle\bar{s}g_s \sigma
Gs\rangle}{32 \pi^2} \int_{\alpha_{i}}^{\alpha_{f}}d\alpha
\alpha(1-\alpha)
\left[6+\left(4s+\frac{s^2}{M^2}\right)\delta(s-\widetilde{\widetilde{m}}^2_Q)
\right] \nonumber  \\
&& +\frac{ m_Q^2\langle\bar{s} s\rangle \langle\bar{s}g_s \sigma
Gs\rangle}{32 \pi^2} \int_{\alpha_{i}}^{\alpha_{f}}d\alpha
\left[1+\frac{s}{M^2}\right] \delta(s-\widetilde{\widetilde{m}}^2_Q)\nonumber  \\
&& -\frac{ 3 \langle\bar{s}g_s \sigma Gs\rangle^2}{128 \pi^2}
\int_{\alpha_{i}}^{\alpha_{f}}d\alpha \alpha(1-\alpha)
\left[1+\frac{s}{M^2}+\frac{s^2}{2M^4}+\frac{s^3}{6M^6}\right] \delta(s-\widetilde{\widetilde{m}}^2_Q)\nonumber  \\
&& -\frac{ 3m_Q^2 \langle\bar{s}g_s \sigma Gs\rangle^2}{768
\pi^2M^6} \int_{\alpha_{i}}^{\alpha_{f}}d\alpha s^2
\delta(s-\widetilde{\widetilde{m}}^2_Q)\, ,
\end{eqnarray}

\begin{eqnarray}
\rho_{\langle GG\rangle}(s)&=& \frac{3}{256 \pi^4}
\int_{\alpha_{i}}^{\alpha_{f}}d\alpha \int_{\beta_{i}}^{1-\alpha}
d\beta \alpha\beta s(s-\widetilde{m}^2_Q)\nonumber  \\
&& +\frac{3}{512 \pi^4} \int_{\alpha_{i}}^{\alpha_{f}}d\alpha
\int_{\beta_{i}}^{1-\alpha}
d\beta (1-\alpha-\beta)^2 s(s-\widetilde{m}^2_Q)\nonumber  \\
&&- \frac{m_Q^2}{1024 \pi^4} \int_{\alpha_{i}}^{\alpha_{f}}d\alpha
\int_{\beta_{i}}^{1-\alpha}
d\beta\left[ \frac{\alpha}{\beta^2}+\frac{\beta}{\alpha^2}\right] (1-\alpha-\beta)^2(3s-2\widetilde{m}^2_Q)\nonumber  \\
&&- \frac{m_Q^4}{1024 \pi^4} \int_{\alpha_{i}}^{\alpha_{f}}d\alpha
\int_{\beta_{i}}^{1-\alpha} d\beta \left[
\frac{1}{\alpha^3}+\frac{1}{\beta^3}\right] (1-\alpha-\beta)^2\nonumber  \\
&& + \frac{3m_Q^2}{1024 \pi^4} \int_{\alpha_{i}}^{\alpha_{f}}d\alpha
\int_{\beta_{i}}^{1-\alpha} d\beta\left[
\frac{1}{\alpha^2}+\frac{1}{\beta^2}\right]
(1-\alpha-\beta)^2(s-\widetilde{m}^2_Q)\, ,
\end{eqnarray}

\begin{eqnarray}
\rho_{\langle
GGG\rangle}(s)&=&\frac{m_Q^4}{8192\pi^6}\int_{\alpha_{i}}^{\alpha_{f}}d\alpha
\int_{\beta_{i}}^{1-\alpha} d\beta (1-\alpha-\beta)^2 \left[
\frac{1}{\alpha^4}+\frac{1}{\beta^4}\right]\delta(s-\widetilde{m}^2_Q)
\nonumber \\
&&-\frac{3m_Q^2}{8192\pi^6}\int_{\alpha_{i}}^{\alpha_{f}}d\alpha
\int_{\beta_{i}}^{1-\alpha} d\beta (1-\alpha-\beta)^2 \left[
\frac{1}{\alpha^3}+\frac{1}{\beta^3}\right] \nonumber \\
&&+\frac{m_Q^2}{8192\pi^6}\int_{\alpha_{i}}^{\alpha_{f}}d\alpha
\int_{\beta_{i}}^{1-\alpha}
d\beta\left[2+\widetilde{m}^2_Q\delta(s-\widetilde{m}^2_Q) \right]\nonumber \\
&&(1-\alpha-\beta)^2\left[
\frac{\alpha}{\beta^3}+\frac{\beta}{\alpha^3}\right]\nonumber \\
&&-\frac{1}{16384\pi^6}\int_{\alpha_{i}}^{\alpha_{f}}d\alpha
\int_{\beta_{i}}^{1-\alpha} d\beta \left[
3s-2\widetilde{m}^2_Q\right](1-\alpha-\beta)^2\left[
\frac{\alpha}{\beta^2}+\frac{\beta}{\alpha^2}\right]\nonumber \\
&&-\frac{m_Q^4}{8192\pi^6}\int_{\alpha_{i}}^{\alpha_{f}}d\alpha
\int_{\beta_{i}}^{1-\alpha} d\beta (1-\alpha-\beta)^2 \left[
\frac{1}{\alpha^3\beta}+\frac{1}{\alpha\beta^3}\right]\delta(s-\widetilde{m}^2_Q)
\nonumber \\
&&+\frac{3m_Q^2}{4096\pi^6}\int_{\alpha_{i}}^{\alpha_{f}}d\alpha
\int_{\beta_{i}}^{1-\alpha} d\beta (1-\alpha-\beta)^2 \left[
\frac{1}{\alpha^2\beta}+\frac{1}{\alpha\beta^2}\right] \nonumber \\
&&-\frac{m_Q^2}{8192\pi^6}\int_{\alpha_{i}}^{\alpha_{f}}d\alpha
\int_{\beta_{i}}^{1-\alpha}
d\beta\left[2+\widetilde{m}^2_Q\delta(s-\widetilde{m}^2_Q) \right]\nonumber \\
&&(1-\alpha-\beta)^2\left[
\frac{1}{\alpha^2}+\frac{1}{\beta^2}\right]\nonumber \\
&&-\frac{3}{16384\pi^6}\int_{\alpha_{i}}^{\alpha_{f}}d\alpha
\int_{\beta_{i}}^{1-\alpha} d\beta \left[
3s-2\widetilde{m}^2_Q\right](1-\alpha-\beta)^2\left[
\frac{1}{\alpha}+\frac{1}{\beta}\right] \, ,
\end{eqnarray}
where $\alpha_{f}=\frac{1+\sqrt{1-4m_Q^2/s}}{2}$,
$\alpha_{i}=\frac{1-\sqrt{1-4m_Q^2/s}}{2}$, $\beta_{i}=\frac{\alpha
m_Q^2}{\alpha s -m_Q^2}$,
$\widetilde{m}_Q^2=\frac{(\alpha+\beta)m_Q^2}{\alpha\beta}$,
$\widetilde{\widetilde{m}}_Q^2=\frac{m_Q^2}{\alpha(1-\alpha)}$, and
$\Delta=4(m_Q+m_s)^2$.

\section*{Acknowledgements}
This  work is supported by National Natural Science Foundation of
China, Grant Number 10775051, and Program for New Century Excellent
Talents in University, Grant Number NCET-07-0282.


\begin{thebibliography}{99}

\bibitem{Yexp}  X. L. Wang  et al,    Phys. Rev. Lett.  {\bf 99}, 142002 (2007).

\bibitem{ISRJpsi}   C. Z. Yuan  et al,   Phys. Rev. Lett.  {\bf 99}, 182004 (2007).

\bibitem{ExpDDbar1}   G. Pakhlova  et al,   Phys. Rev. Lett.  {\bf 98}, 092001 (2007).


\bibitem{ExpDDbar2} G. Pakhlova  et al,   Phys. Rev. Lett.  {\bf 100}, 062001  (2008).


\bibitem{ExpDDbar3} G. Pakhlova  et al,   Phys. Rev.  D {\bf 77}, 011103   (2008).

\bibitem{ExpDDbar4} B. Aubert  et al,  arXiv:0710.1371.


\bibitem{Abe:2007sy}   P. Pakhlov  et al, Phys. Rev. Lett. {\bf 100}, 202001
(2008).


\bibitem{Ding0708}   G. J. Ding, J. J. Zhu and M. L. Yan,
  Phys. Rev.  D {\bf 77}, 014033 (2008).

  \bibitem{Chao0903} B. Q. Li and K. T. Chao, Phys. Rev. {\bf D79} (2009) 094004.


\bibitem{SDmix}  A. M. Badalian, B. L. G. Bakker and I. V. Danilkin,
Phys. Atom. Nucl. {\bf 72} (2009) 638.

\bibitem{Qiao0709}   C. F. Qiao,  J. Phys. {\bf G35}, 075008 (2008).

\bibitem{Nielsen0804} R. M. Albuquerque and M. Nielsen, Nucl. Phys. A {\bf 815}, 53 (2009).

\bibitem{babarconf0801}  R. Faccini, arXiv:0801.2679.

\bibitem{Voloshin0803} S. Dubynskiy and M. B. Voloshin, Phys. Lett. B {\bf 666}, 344 (2008).


\bibitem{VoloshinReV} M. B. Voloshin, Prog. Part. Nucl. Phys. {\bf 61} (2008) 455.



\bibitem{Guo0803} F. K. Guo, C. Hanhart and Ulf-G. Meissner, Phys. Lett. B {\bf 665}, 26 (2008).

\bibitem{PDG} C. Amsler et al, Phys. Lett. {\bf  B667},  1 (2008).

\bibitem{GuoPRL} F. K. Guo, C. Hanhart and U. G. Meissner, Phys. Rev. Lett. {\bf 102} (2009) 242004.

\bibitem{WangZhang} Z. G. Wang and X. H. Zhang, arXiv:0905.3784.

\bibitem{SVZ79}  M. A. Shifman, A. I. Vainshtein and V. I. Zakharov,
Nucl. Phys. {\bf B147} (1979) 385, 448.

\bibitem{Reinders85} L. J. Reinders, H. Rubinstein and S. Yazaki, Phys. Rept. {\bf 127} (1985) 1.


\bibitem{Wang2004} Z. G. Wang, W. M. Yang and S. L. Wan, Eur. Phys. J.  {\bf C37}, 223 (2004).
\bibitem{Close2002} F. E. Close and N. A. Tornqvist, J. Phys.  {\bf G28} (2002) R249.

\bibitem{ReviewScalar}   C. Amsler and N. A. Tornqvist, Phys. Rept. {\bf 389} (2004) 61.

\bibitem{Lee2005} S. H. Lee, H. Kim and  Y. Kwon, Phys. Lett. {\bf B609} (2005)
252.

\bibitem{Nielsen0907} R. D. Matheus, F. S. Navarra, M. Nielsen and C. M.
Zanetti, Phys. Rev. {\bf D80} (2009) 056002.

\bibitem{Ioffe2005} B. L. Ioffe, Prog. Part. Nucl. Phys. {\bf 56} (2006)
232.


\bibitem{Wang1} Z. G. Wang, Nucl. Phys. {\bf A791} (2007) 106.

\bibitem{Wang2} Z. G. Wang, W. M. Yang and S. L. Wan, J. Phys. {\bf G31} (2005) 971.

\bibitem{Wang08072} Z. G. Wang, Eur. Phys. J. {\bf C62} (2009) 375.


\bibitem{Wang0904} Z. G. Wang, Eur. Phys. J. {\bf C63} (2009) 115.


\bibitem{LHC}  G. Kane and A. Pierce, "Perspectives On LHC Physics",
World Scientific Publishing Company,  2008.



\end{thebibliography}
\end{document}